\documentclass[prd,superscriptaddress,nofootinbib]{revtex4}

\usepackage{framed}
\usepackage{bbold}
\usepackage{slashed}
\usepackage{graphicx}
\usepackage{amsmath}
\usepackage{amssymb}
\usepackage{epsfig}
\usepackage{hhline}
\usepackage{soul}
\usepackage{tabularx,pifont,multirow}
\usepackage{enumitem}
\usepackage{colordvi}
\usepackage{soul}
\usepackage{xcolor}
\usepackage{yfonts}

\newcommand{\be}{\begin{equation}}
\newcommand{\ee}{\end{equation}}

\definecolor{darkgreen}{rgb}{0,0.3,0.05}
\makeatletter                                                         %
\newcommand*\rel@kern[1]{\kern#1\dimexpr\macc@kerna}                  %
\newcommand*\widebar[1]{                                              %
  \begingroup                                                         %
  \def\mathaccent##1##2{                                              %
    \rel@kern{0.8}                                                    %
    \overline{\rel@kern{-0.8}\macc@nucleus\rel@kern{0.2}}             %
    \rel@kern{-0.2}                                                   %
  }                                                                   %
  \macc@depth\@ne                                                     %
  \let\math@bgroup\@empty \let\math@egroup\macc@set@skewchar          %
  \mathsurround\z@ \frozen@everymath{\mathgroup\macc@group\relax}     %
  \macc@set@skewchar\relax                                            %
  \let\mathaccentV\macc@nested@a                                      %
  \macc@nested@a\relax111{#1}                                         %
  \endgroup                                                           %
}                                                                     %
\makeatother                                                          %

\begin{document}

\preprint[\leftline{KCL-PH-TH/2017-{\bf 62}, \,
IFIC/17-59}

%

\title{\Large {\bf Leptogenesis from Heavy Right-Handed Neutrinos in CPT Violating Backgrounds } }

\bigskip

\author{Thomas Bossingham}

\affiliation{\vspace{1mm}
Theoretical Particle Physics and Cosmology Group,
  Department of Physics, King's College London, Strand, London WC2R
  2LS, UK}

\author{Nick E.~Mavromatos}

\affiliation{\vspace{1mm} Theoretical Particle Physics and Cosmology Group, Department of Physics, King's College London,  
Strand, London WC2R 2LS, UK}
\affiliation{\vspace{1mm} Departament de F\'isica Te\`orica and IFIC, Universitat de Val\`encia - CSIC, E-46100, Spain}
  
\author{Sarben Sarkar}

\affiliation{\vspace{1mm}
Theoretical Particle Physics and Cosmology Group,
  Department of Physics, King's College London, Strand, London WC2R
  2LS, UK}


\begin{abstract}
\vspace{0.5cm}
\centerline{\bf Abstract }
\noindent\\[-2mm] 
We discuss leptogenesis in a model with heavy right-handed Majorana neutrinos propagating in a constant but otherwise generic CPT-violating axial time-like background (which could  be motivated by string theory considerations). At temperatures much higher than the temperature of the  electroweak phase transition we solve analytically but approximately (using Pad\'e approximants) the corresponding Boltzmann equations, which 
describe lepton asymmetry generation due to the tree-level decays of the heavy neutrinos into standard model leptons. These leptons are effectively massless at such temperatures. The current work completes in a rigorous way a preliminary treatment of the same system, by some of the present authors. In this earlier work, lepton asymmetry was crudely estimated considering the decay of a right-handed neutrino at rest.  Our present analysis includes thermal momentum modes for the heavy neutrino and this leads to a total lepton asymmetry which is bigger by a factor of two as compared to the previous estimate. Nevertheless, our current and preliminary results for the freezeout are found to be in agreement (within a $\sim 12.5\%$  uncertainty). Our analysis depends on a novel use of Pad\'e approximants to solve the Boltzmann equations and may be more widely useful in cosmology.

\end{abstract}
\maketitle

\section{Introduction and Motivation}

A plethora of cosmological measurements, especially those associated with observations of the Cosmic Microwave background Radiation (CMB) in the Universe~\cite{Planck}, 
estimates the 
observed asymmetry between matter (mostly baryons) and antimatter to be of order:
\begin{equation}
\Delta n(T\sim 1~{\rm GeV})=\frac{n_{B}-n_{\overline{B}}}{n_{B}+n_{\overline{B}}}\sim\frac{n_{B}-n_{\overline{B}}}{s}=(8.4-8.9)\times10^{-11}
\label{basym}
\end{equation}
at the early stages of the cosmic expansion, \emph{i.e.} for times $t<10^{-6}$~sec
and temperatures $T>1$~GeV. In the above formula $n_{B}$ ($n_{\overline{B}}$)
denotes the (anti) baryon density in the universe, and $s$ is the
entropy density of the Universe. Moreover, the observations indicate that at present, where the temperature of the Universe is that of the 
CMB background, $T_0 = 2.727\, {\rm K} = 0.235 \, {\rm meV}$, the ratio of baryons over  photons  is  
\begin{equation}\label{photonbaryon}
\frac{n_B}{n_\gamma} \sim 5.4 \, \times 10^{-10}~, 
\end{equation}
where $n_\gamma$ is the density of photons in the Universe.

At first sight, the asymmetry (\ref{basym}) (and the result (\ref{photonbaryon})) appears to be in conflict with fundamental properties of relativistic quantum field theories, on which we base our phenomenology of elementary particles.
Specifically, any Lorentz invariant quantum field theory, formulated on a flat space-time, which respects unitarity and locality, should be described by a Lagrangian that is 
invariant under CPT transformations (at any permutation of the operations), where C denotes Charge conjugation, T reversal in time and P parity (spatial reflexion) transformations. 
This is the celebrated \emph{CPT theorem}~\cite{cpttheorem}. 
For the physics of the  the early universe based on any Lorentz invariant quantum field theory, such a theorem implies that matter and antimatter should be created in equal amounts after the Big Bang. If such is the case, the universe today would be filled with radiation, as a result of matter-antimatter annihilation processes, in conflict with (\ref{photonbaryon}). 

Within the context of our current understanding of fundamental physics, A. Sakharov~\cite{Sakharov}, postulated the following three necessary conditions for the dominance of matter over antimatter (baryon asymmetry in the universe (BAU) (\ref{basym})), and hence for our very existence today: 
\begin{itemize}
\item
Baryon (B) number violation.
\item
Charge (C) and Charge-Parity (CP) symmetries need to be broken.
\item
Chemical equilibrium does not hold during an epoch in the early universe, since chemical equilibrium washes out asymmetries.
\end{itemize}
In fact there are two types of non-equilibrium processes
in the early universe that can  produce asymmetries between particles and antiparticles: the first
type concerns processes generating asymmetries between leptons and
antileptons (\emph{leptogenesis})~\cite{Buchmuller:2004nz,Davidson:2008bu,Pilaftsis:2013nqa}, while the second produces asymmetries
between baryons and antibaryons directly (\emph{baryogenesis})~\cite{Cohen:1993nk,Trodden:1998ym,Riotto:1999yt,Buchmuller review}. 

Unfortunately, within the Standard Model (SM) framework, although Sakharov's axioms can be qualitatively reproduced, especially because one has both B and CP violation in the quark sector, 
the resulting baryon asymmetry is several orders of magnitude 
smaller than the observed one (\ref{basym})~\cite{Kuzmin:1985mm,Gavela:1993ts,Gavela:1994dt}.
There are several ideas that go beyond the SM (\emph{e.g}.  grand unified
theories, supersymmetry, extra dimensional models \emph{etc.}) and provide extra sources of CP violation, necessary for yielding the observed magnitude for the  asymmetry. 
Some of these attempts, involve the elegant mechanism of baryogenesis via leptogenesis, in  which a lepton asymmetry is generated first, by means of 
decays of right handed sterile neutrinos to SM particles; the lepton asymmetry is subsequently communicated to the baryon sector by means of sphaleron processes which 
violate both Baryon (B) and Lepton (L) numbers, but preserve the difference B-L~\cite{Fukugita-Yanagida,Luty,Pilaftsis:1997jf,Buchmuller:2005eh,strumia,Shaposhnikov:2006xi}.
Heavy sterile neutrinos, through the the seesaw mechanism~\cite{seesaw}, play another essential r\^ole in particle physics, since they provide a natural explanation for the existence of three light neutrinos with masses small compared to other mass scales in the SM), as suggested by observed neutrino oscillations~\cite{oscil}.
Fine tuning and some \emph{ad hoc} assumptions are involved though in such scenarios, especially in connection with the magnitude of the CP violating phases and the associated decay widths. Consequently the quest for a proper understanding of the observed BAU still requires
further investigation.

In the scenario of Sakharov it is assumed that \emph{CPT} 
symmetry holds in the very early universe and this leads to the equal production
of matter and antimatter. \emph{CPT} invariance is regarded as fundamental
since it is a direct consequence of the celebrated \emph{CPT} theorem~\cite{cpttheorem}.
However, it is possible that some of the assumptions in the proof of  the \emph{CPT}  theorem do not hold in the early universe, 
leading to violations of \emph{CPT}  symmetry. Sakharov has stated that non-equilibrium processes are necessary for BAU in \emph{CPT}  invariant theories. If the
requirement of \emph{CPT} is relaxed,  the necessity
of non-equilibrium processes can be dropped . In a low-energy version of quantum gravity Lorentz invariance and unitarity are likely to emerge since not all degrees of freedom are accessible to a low-energy observer. Lorentz invariance violation has been singled out in ref.~\cite{Greenberg} as a fundamental reason for inducing \emph{CPT}  violation (CPTV) and vice versa. (However, such claims have been disputed in \cite{Chaichian}, through counterexamples of Lorentz invariant systems, which  
violate \emph{CPT}  through relaxation, for example, of locality.) In our work we will consider Lorentz invariance violating (LV) backgrounds in the early universe as a form of \emph{spontaneous} violation of Lorentz  and \emph{CPT}  symmetry. 

If LV is the primary source of CPTV, then the latter can be studied within a local effective field theory framework, which is known as the \emph{Standard Model Extension}(SME)~\cite{sme}. The latter provides the most general parametrization for studying the phenomenology of Lorentz violation in a plethora of physical systems, ranging from cosmological probes, to particle and precision atomic physics systems. For the current era of the universe~\cite{smebounds} very stringent upper bounds on the potential amount of Lorentz and \emph{CPT}  violation have been placed by such systems. 
However, under the extreme conditions present in the very early universe, such violations could be significantly stronger than in the present era (where they could be extremely suppressed (or absent), in agreement with current stringent constraints).~\footnote{If one considers, for instance, quark fields in some Lorentz and CPTV backgrounds (such as those allowed by the SME formalism), it is possible to induce baryogenesis, as a consequence of the fact that the LV and CPTV effects induce ``chemical potentials'' for the quarks~\cite{Bertolami}. This leads directly to baryogenesis, given that in the presence of a chemical potential $\mu$, the populations of quarks and antiquarks are already different within thermal equilibrium, since the 
the particle  and antiparticle phase-space distribution functions $f(E,\mu), f(\overline E, \overline \mu)$, with $E$ the energy (and an overline over a quantity denoting that of an antiparticle)
are different (in the presence of a chemical potential, $\mu$, for a particle, the antiparticle has a chemical potential of opposite sign $\overline \mu = -\mu$. In SME models, of course, even the magnitudes of $\overline \mu$ and $\overline E$ may be different from those of particles, as a consequence, for example, of different dispersion relations between particles and antiparticles). All these cause a difference in the corresponding equilibrium populations
\begin{eqnarray}~\label{cptvf}
f(E,\mu)=[{\rm exp}(E-\mu)/T)\pm1]^{-1}~, \quad 
f(\overline{E},\bar{\mu})=[{\rm exp}(\bar{E}-\overline{\mu})/T)\pm1]^{-1}~,
\end{eqnarray}
(where the $+ (-)$ will denote a fermionic (bosonic) (anti-)particle). In principle, such scenarios can lead to alternative explanations for the observed matter-antimatter asymmetry, provided that detailed mechanisms for freeze-out of particle interactions in this SME context are provided.Unfortunately, so far, microscopic models leading to such SME lagrangians and related phenomena have not been provided. }
In a  previous work~\cite{decesare} we presented a phenomenological model for generating a lepton asymmetry via CPTV in the early universe. 
The model was based on a specific extension of the SM, involving massive Majorana right-handed neutrinos (RHN), propagating on a Lorentz and CPTV, constant in time, axial vector background coupling to fermions. The latter could be traced back to a specific configuration of a cosmological Kalb-Ramond antisymmetric 
tensor field~\cite{kalb} that appears in the gravitational multiplet of string theory~\cite{sloan,kaloper,aben}, and plays the r\^ole of torsion in a generalised connection, although such an identification is not restrictive. The involvement of sterile 
RHN in the model is physically motivated primarily by the need to provide a natural explanation for the light neutrino masses of the SM sector. 
The lightest RHN may also have a potential role as (warm) dark matter candidates~\cite{Shaposhnikov:2006xi,neutrinoDM}. However, in our CPTV models sterile neutrinos responsible for leptogenesis have masses in the $10^5$ GeV range or higher~\cite{decesare}) and so cannot be considered as dark matter. 

In \cite{decesare} we only gave a qualitative and rather crude estimate of the induced CPTV lepton asymmetry, based on the decaying right handed Majorana 
neutrino being at rest. In this way it was possible to estimate the lepton asymmetry, without following the standard procedure of solving the appropriate Boltzmann equation that determines correctly the asymmetry value at decoupling of RHN. In the early universe the heavy right-handed neutrinos are not at rest but have a Maxwell-Boltzmann momentum distribution. The purpose of this article is to properly take into account this momentum distribution in the calculation of the lepton asymmetry.

 The structure of the article is as follows: in the next section \ref{sec:review} we review the model of \cite{decesare} and an earlier estimate of the CPTV-background induced lepton asymmetry, which shall be compared with the much more accurate result  of the present article, obtained by solving the appropriate Boltzmann equations analytically.  In section \ref{sec:boltz}, 
 we construct the appropriate system of Boltzmann equations in the presence of a weak CPTV axial background involved in the problem, and compare it with the standard CP violating case~\cite{Fukugita-Yanagida,Luty,Pilaftsis:1997jf,Buchmuller:2005eh,strumia,Shaposhnikov:2006xi}.
In section \ref{sec:sol}, we solve the Boltzmann equations using Pad\' e approximants~\cite{pade}, which is an approximation popular in several fields of physics, ranging from statistical mechanics to particle physics and quantum field theory~\cite{padeparticle}. In this way, we manage to compute the induced lepton asymmetry at RHN decoupling analytically, avoiding numerical treatment. It should be remarked,  that 
setting up and solving such a system of differential equations is a highly non-trivial and algebraically complicated task. Our analytical results agree (within $\sim 12.5\%$ accuracy) with our earlier preliminary estimates of the freezeout point, as outlined, in \cite{decesare}. In view of this, we consider our system of Boltzmann  equations as providing another efficient use of Pad\`e approximants, this time with relevance to cosmology. The lepton asymmetry that we find in our analytic treatment is slightly larger (by a factor of about 2) than the estimate of \cite{decesare}; this is to be expected, since non-zero momentum modes of the RHN have been included. 
Conclusions and outlook are given in section \ref{sec:concl}. A review of the formalism and derivations of the corresponding decay amplitudes and thermally averaged rates used in the Boltzmann equations,  are presented in several Appendices.  

\section{Review of the CPT Violating Model for Leptogenesis \label{sec:review}}

It will suffice for our purposes to consider a single species of RHN as in \cite{decesare}. If the phenomenology is required to include the seesaw mechanism it is necessary (and possible) to add more species of RHN.  The option of using a single species of RHN is not available within the standard \emph{CPT} conserving but CP violating scenario, where to obtain a lepton asymmetry one needs more than one species of RHN~\cite{Fukugita-Yanagida,Luty,strumia}. Our Lagrangian is given by~\cite{decesare}:
\be
\label{smelag}
\mathcal{L}=i\overline{N}\slashed{\partial}N-\frac{M}{2}(\overline{N^{c}}N+\overline{N}N^{c})-\overline{N}\slashed{B}\gamma^{5}N-y_{k}\overline{L}_{k}\tilde{\phi}N+h.c.
\ee
where 
$N$ is the Majorana field,  $\tilde \phi$ is the adjoint ($\tilde{\phi}_i=\varepsilon_{ij}\phi_j $) of the Higgs field  $\phi$, 
 and $L_{k}$ is a lepton (doublet) field of the SM sector, with $k$ a generation index. $y_k$ is a Yukawa coupling, which is non-zero and provides a non-trivial interaction between the RHN and the SM sectors via the Yukawa type interaction (``Higgs portal''):
$\mathcal{L}_{YUK} = y_{k}\overline{L}_{k}\tilde{\phi}N+h.c$. In our case of a single Majorana neutrino species we take $k=1$ to label the first generation, and from now on we set 
\begin{equation}\label{yc1}
y_1 = y ~.
\end{equation}
Since in SM the leptons have definite chirality, the Yukawa interactions  $\mathcal{L}_{YUK}$ can be rewritten as
\be\label{yuk}
\mathcal{L}_{YUK}=-y \overline{L}_{1}\tilde{\phi}\left(\frac{1+\gamma^{5}}{2}\right)N-y^{*}\overline{N}\tilde{\phi}^{\dagger}\left(\frac{1-\gamma^{5}}{2}\right)L_{1} =
-y \overline{L}_{1}\tilde{\phi}\left(\frac{1+\gamma^{5}}{2}\right)N-y ^{*}\overline{L}_{1}^{c}\tilde{\phi}^{\dagger}\left(\frac{1-\gamma^{5}}{2}\right)N.
\ee
where in the last equality we used the properties of the charge conjugation matrix and the Majorana condition $N^c=N$.
The two hermitian conjugate terms in the Yukawa Lagrangian are also \emph{CPT} conjugate. This is to be expected on the basis of the \emph{CPT} theorem. In fact \emph{CPT} violation is introduced only by interactions with the background field.

The background field $\slashed{B} \equiv \gamma_\mu \, B^\mu$ is assumed at most a function of the cosmic time, so as to respect the isotropy and homogeneity of the early universe, where such backgrounds are non-trivial. We note at this point that, if the the axial background field $B^\mu$ is to be identified~\cite{decesare} with the totally antisymmetric field strength ($H_{\mu\nu\rho} = \partial_{\mu} B_{\rho\sigma}$ +  cyclic permutation of indices) of the Kalb-Ramond~\cite{kalb} spin-one field $B_{\mu\nu}$, that appears in the massless gravitational multiplet of string theory~\cite{sloan}, then the latter is viewed as 
part of a torsion
background~\cite{kaloper}: $B^\mu = \epsilon^{\mu\nu\rho\lambda}\, H_{\nu\rho\lambda}$. In such a case one should also consider 
the coupling of the axial field $B_\mu$ to all other fermions of the SM sector, $\psi_j$ ($j$=leptons, quarks) via a \emph{universal} minimal prescription,
with the coupling with all fermionic species $\psi$ being the same : $\overline \psi_j\,\gamma^5 \, \slashed{B}\, \psi_j $. 
In four space-time dimensions the $H_{\nu\rho\lambda}$ field 
is dual to a pseudoscalar field $b(x)$~\cite{aben,kaloper}: $H_{\mu\nu\rho} \propto \epsilon_{\mu\nu\rho\lambda}\, \partial^\lambda b$. There is an exact cosmological solution 
in  the bosonic string theory~\cite{aben}, in which the $H$-torsion 
background is identified with a homogeneous and isotropic cosmological Kalb-Ramond axion, linearly dependent on the cosmic time~\cite{aben}. The solution satisfies the corresponding conformal invariance conditions of the associated $\sigma$-model, thus constituting a consistent background of strings. 
The resultant axial backgrounds are constant in time and have non-trivial temporal components only 
\begin{equation}\label{temporalB}
B_0 = {\rm const} \ne 0~, \, B_i = 0 ~, i=1,2,3~.
\end{equation}
In \cite{decesare} we have generalised the above solution (\ref{temporalB}) in theories with fermions, in which the latter condensed in the early universe.
Such backgrounds can then be viewed as spontaneously breaking Lorentz and \emph{CPT} symmetry in the system and are consistent with isotropy and homogeneity of the early universe.
In what follows we shall consider the Lagrangian (\ref{smelag}) in the generic background (\ref{temporalB}), without specifying further its microscopic origin. The form of the Lagrangian 
coincides with one of the simplest forms of the so-called Standard Model Extension (SME)~\cite{sme}, namely that in which the temporal component of the so-called $b_\mu$ coefficient assumes a constant value. 

There are stringent constraints~\cite{smebounds} (coming from a plethora of measurements ranging from astrophysical to laboratory precision tests of Lorentz and CPT symmetries) for today's value of $b_0 \le 0.02 $ eV (and much suppressed spatial components $b_i  < 10^{-32}$ GeV). Although in  our model in the frame of Robertson-Walker (Cosmic Microwave Background) the axial background is assumed to have only the temporal component (\ref{temporalB}), nevertheless the slightest motion of the observer with respect to that frame will generate a spatial component by means of a Lorentz transformation. It is therefore essential that any current value of $B_0$ is severely suppressed today, and also during the nucleosynthesis era. In \cite{decesare} we have provided arguments in favour of scenarios in which the universe undergoes a phase transition soon after the decoupling of 
 heavy neutrinos, so that the background $B_0$ ceases to be a constant, and decreases with the temperature according to the scaling law $T^3$. The qualitative estimates of \cite{decesare}, have indicated that for Yukawa couplings $y_k $ of order $10^{-5}$ (assumed in \cite{decesare}), the decoupling temperature of the heavy neutrino  $T_D$ of order $T_D \simeq m_N \sim 100$~TeV, implies a phenomenologically consistent leptogenesis for $B_0 \sim 1$ MeV at $T\simeq T_D$. Soon after, the cooling law $B_0 \sim T^3$ implies for the present era a negligible $B_0 = \mathcal{O}(10^{-44})$~meV today, and also a very small value during the nucleosynthesis era.

\begin{figure}
\includegraphics[width=0.4\columnwidth]{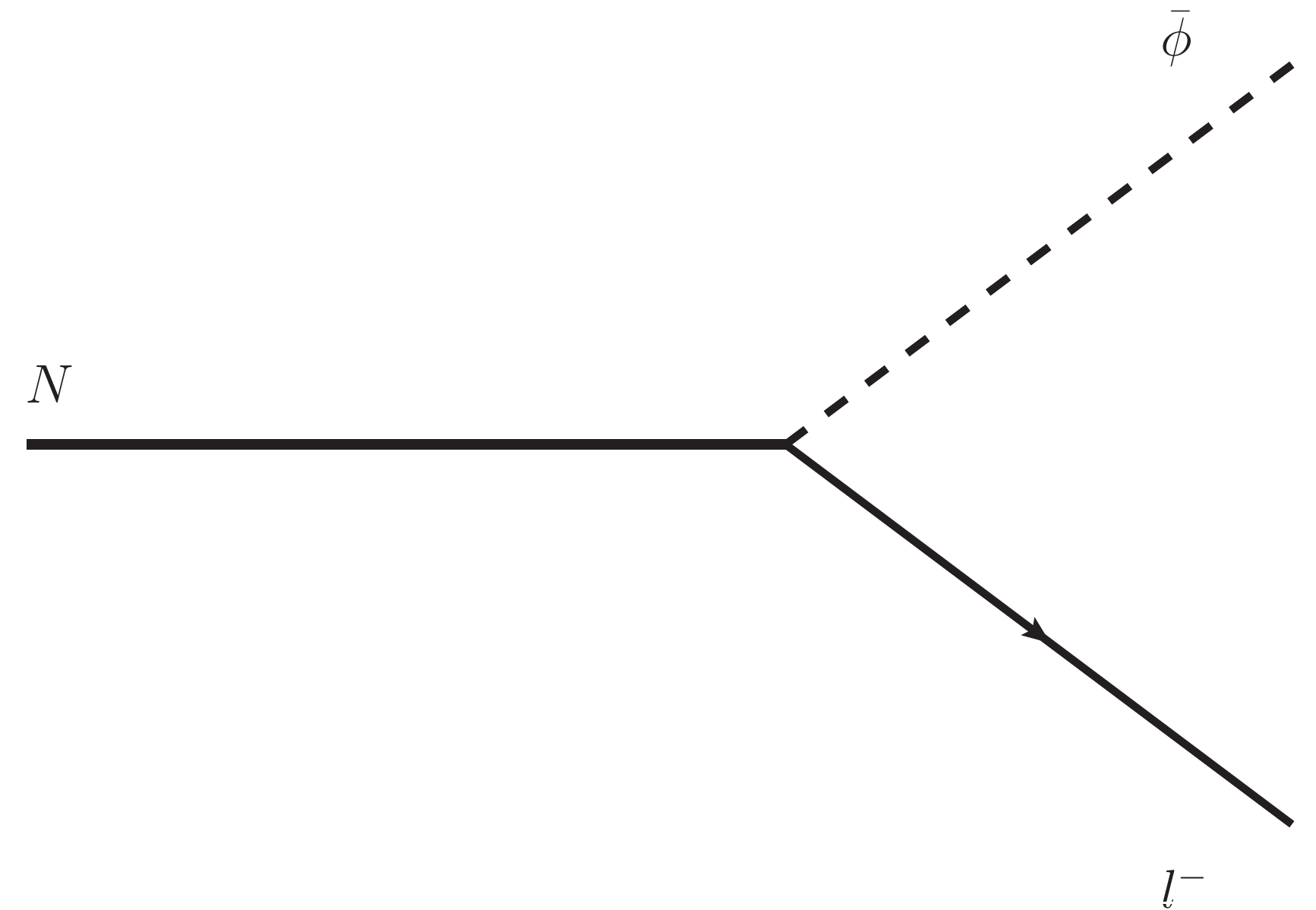} \hspace{2cm}
\includegraphics[width=0.4\columnwidth]{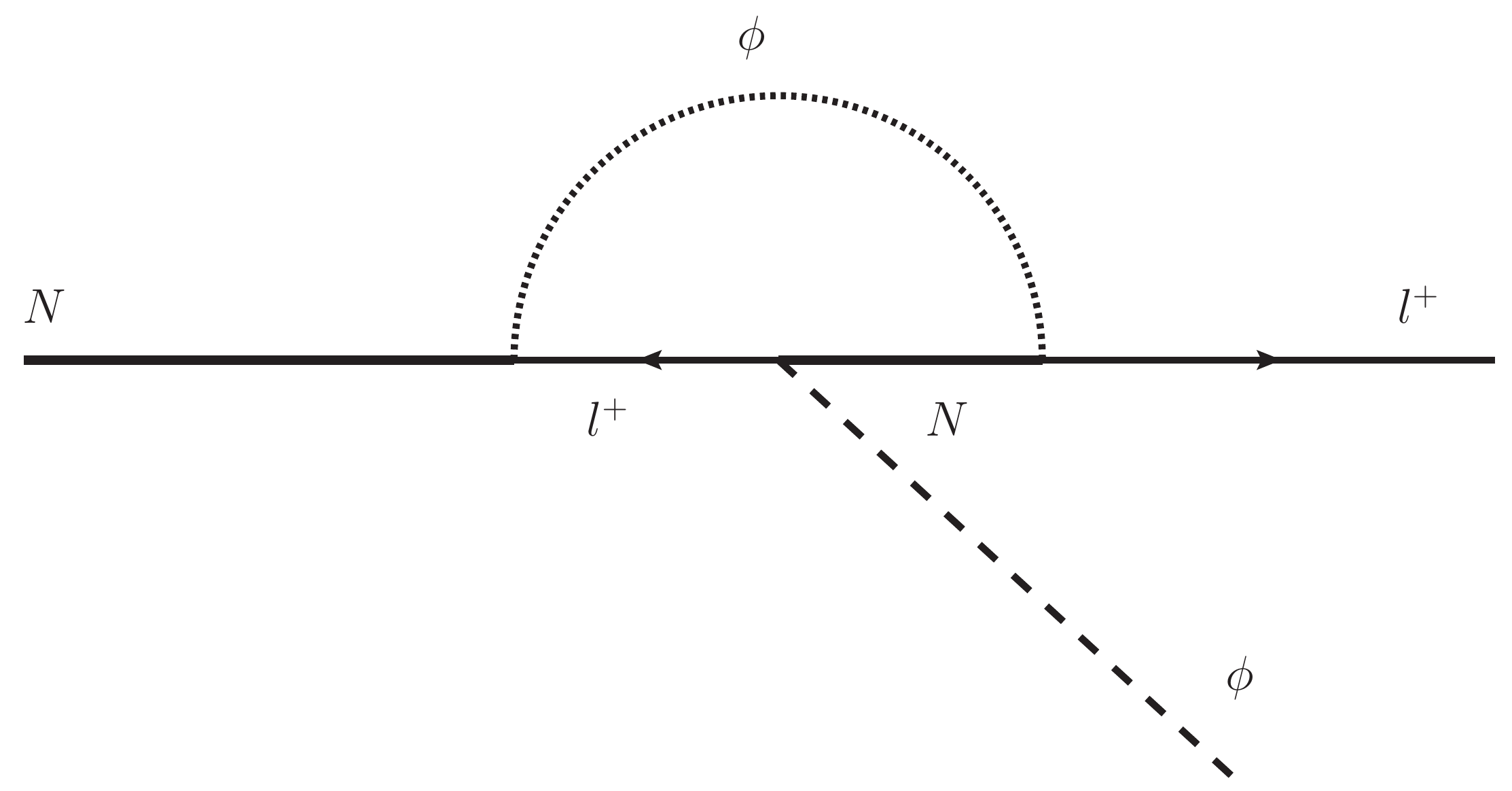} 
\caption{Tree- (left) and one-loop (right) decay amplitudes for the decays (\ref{2channels}) that are relevant for leptogenesis. Continuous undirected lines represent right-handed neutrinos, lines with an arrow represent SM leptons,  whilst dashed lines correspond to the SM Higgs. In our approach, the left diagrams are evaluated in the presence of an axial background field (\ref{temporalB}). The right diagram is the standard result of \cite{Fukugita-Yanagida}, leading to Leptogenesis in CPT invariant theories, with only CP violation in the lepton sector.}
\label{fig:decays}
\end{figure}

As we shall be interested in high temperatures $T \simeq T_D \sim 100$~TeV, which are much higher than the electroweak phase transition, the SM fields are treated as massless, while the heavy RHN can still be assumed to be massive~\footnote{We do not specify here or in \cite{decesare} the mechanism by which the heavy right-handed neutrinos acquire their mass. Exotic scenarios may be at play here~\cite{MavPilaftsis}, in which the quantum fluctuations of the Kalb-Ramond $H_{\mu\nu\rho}$ field (equivalently the axion field $b(x)$ in four space-time dimensions) are allowed to mix with ordinary axions, via kinetic mixing, and thus may be responsible for radiative generation of the right-handed Majorana neutrino mass, as a result of Yukawa coupling interactions of the ordinary axion with such right-handed neutrinos. In such a case, one may arrange that such masses are non trivial in the high temperature regime of the decoupling of the right-handed neutrinos, even if the rest of the SM fields are massless at such temperatures.}. In such a case, the Higgs field does not develop a vacuum expectation value; consequently the charged  Higgs (denoted by $h^\pm$) and neutral Higgs ($h^0$) play a r\^ole in the physical spectrum. From the form of the interaction Lagrangian in Eqns. (\ref{smelag}), and (\ref{yuk}), it is straightforward to obtain the Feynman rules for the diagrams giving the decay of the Majorana particle in the two distinct channels:
\begin{eqnarray}\label{2channels}
{\rm Channel ~I}&:& \qquad  N \rightarrow l^{-}h^{+}~,  \\ \nonumber 
{\rm Channel ~II}&:& \qquad  N \rightarrow l^{+}h^{-}~.
\end{eqnarray}
The neutral channel decay $N \to \nu \, h^0$, where $\nu$ are the SM sector neutrinos, does not lead to any lepton asymmetry, as follows directly from the Yukawa term (\ref{yuk}), when  expressed in terms of Majorana fields for the neutrinos. In the absence of the background, the squared matrix elements obtained from tree level diagrams for the two decays (\ref{2channels}) (\emph{cf.} figure \ref{fig:decays}) would be the same \cite{Kolb Wolfram,Fukugita-Yanagida,Luty,strumia}. In such a case, a lepton asymmetry is generated due to the CP violation present in the one loop diagram. In the presence of the background $B_0 \ne 0$, however, there is a difference in the decay rates of the tree level processes (\ref{2channels}), and this leads to CPTV-induced lepton asymmetry~\footnote{Scattering processes $l \, l \rightarrow \bar h \bar h$ or $l \, h \rightarrow \bar l \, \bar h $, are of higher order in the Yukawa coupling $y$ and hence are suppressed in our case, although such processes are equally important in standard CPT invariant, CP violating leptogenesis, with more than one species of right-handed neutrinos, as they are of the same order as the CP violating one-loop graphs~\cite{strumia}.}.
 In \cite{decesare}, by assuming the heavy Majorana neutrino at rest, we estimated the lepton asymmetry induced by the (Lorentz-and-CPT-violating) background $B_0$. We assumed one single Majorana neutrino $N$ with the corresponding Yukawa coupling for the Higgs portal $y$. 
For $N$, the \emph{tree-level} decays (\emph{cf.} fig.~\ref{fig:decays}) for the two channels (\ref{2channels}), in the presence of the background $B_0$, yields in that case:
\be\label{sdecayrates}
\Gamma_1 (N \rightarrow l^{-}h^{+})= \frac{|y|^2}{32\pi^2}\frac{m_N^2}{\Omega}\frac{\Omega+B_0}{\Omega-B_0}, \qquad 
\Gamma_2 (N\rightarrow l^{+}h^-) = \frac{|y|^2}{32\pi^2}\frac{m_N^2}{\Omega}\frac{\Omega-B_0}{\Omega+B_0}~, \qquad \Omega=\sqrt{B_0^2+m_N^2}~.
 \ee
The decay process goes out of equilibrium when the total decay rate drops below the expansion rate of the universe. Assuming standard cosmology~\cite{decesare} during the decoupling period~\footnote{Such an assumption is non trivial and depends on the microscopic model considered. For instance, in terms of brane-world scenarios  for the background $B_0$~\cite{decesare}, where the latter is derived from a cosmological Kalb-Ramond axion field $b(t)$, such an assumption is justified by requiring a cancellation of the constant in time kinetic energy density of the field $b$ by the (negative) dark energy of the higher-dimensional bulk. After the decoupling, where the string/brane Universe undergoes a phase transition, the dark energy falls off with the temperature sufficiently rapidly, so as today it reaches the value measured by cosmological observations. We shall not discuss such details in the current article.}, which is also hypothesised to coincide with the radiation-dominated era of the Universe, this expansion rate is given by the Hubble constant~\cite{Weinberg}
\be
\Gamma\simeq H=1.66 \, T^2 \mathcal{N}^{1/2} M_{pl}^{-1}~,
\ee
where $\mathcal{N}$ is the effective number of degrees of freedom of all elementary particles and $M_{pl}$ is the Planck mass. For a minimal extension of the SM, with only right-handed neutrinos and the background $B_0$, we may estimate $\mathcal{N} = \mathcal{O}(100)$ at temperatures higher then the electroweak transition~\cite{Kolb}.
From the last equation one can estimate the right-handed-neutrino decoupling temperature $T_{D}$, in terms of the phenomenological parameters $\Omega$, $|y|$ and $B_0$~\cite{decesare}
\be\label{freeze}
T_{D}\simeq6.2\cdot 10^{-2} \frac{|y|}{\mathcal{N}^{1/4}}\sqrt{\frac{M_{pl}(\Omega^2+B_0^2)}{\Omega}}.
\ee
Imposing a delayed decay mechanism, as for the standard leptogenesis~\cite{Weinberg, Weinberg 2, Fukugita-Yanagida}, leads to the further requirement that $T_{D}\leq \Omega$ 
leading to:
$\xi \, (\Omega^2+B_0^2)\leq\Omega^3$, where $\xi =3.8\cdot 10^{-3}\frac{m_{P}|y|^2}{\mathcal{N}^{1/2}}$. In \cite{decesare} we demanded that saturation of this inequality be satisfied for all values of the background field $B_0$, which implies
\be\label{lowerb}
m_N^2\geq 1.09\, \xi^2.
\ee
On assuming for the (phenomenological) coupling $y$ the value $|y|\approx 10^{-5}$, we then obtain an order of magnitude estimate  $\overline{m_N}$ for the heavy neutrino mass 
\be\label{hnmass}
\overline{m_N}\approx  T_D \approx 100 \;\mbox{TeV}~. 
 \ee
In \cite{decesare} we estimated the lepton number density by assuming that all the right-handed neutrinos were at rest before the decay; hence with branching ratios of the decays  given by $r=\frac{\Gamma_1}{\Gamma}$ and $1-r$, the decay of a single neutrino produces the lepton number
\be \label{lepton number production in one decay}
\Delta L=r-(1-r)=2r-1=\frac{2\Omega B_0}{\Omega^2+B_0^2}.
\ee
Multiplying this quantity by the initial abundance of right-handed Majorana neutrinos $N_D$ at the temperature $T_D$ (averaged over the respective helicities), one gets a $crude$ estimate of the lepton number density. Also, in \cite{decesare} we assumed that the right-handed neutrino density distribution follows closely the equilibrium distribution for $T\geq T_{D}$ and drops rapidly to zero at lower temperatures $T\leq T_{D}$; furthermore  the density of the sterile neutrino (normalised to the entropy density) is well approximated by a step-function. This implies that the total lepton asymmetry (normalised over the entropy density) produced in the full decay of the right-handed neutrino is given by~\cite{decesare}
\be\label{dl}
\frac{\Delta L^{TOT}}{s} (T \simeq T_D)=(2r-1)\frac{\bar n_{N}}{s}=\frac{2\Omega B_0}{\Omega^2+B_0^2} \frac{\bar n_N^{eq}}{s}
\ee
where 
\be\label{entropydensity}
s \sim \frac{2\pi}{45} {\mathcal N} \, T^3 \sim 14 T^3~,
\ee is the total entropy density (assuming, for temperatures higher than the electroweak phase transition, SM-like values for the effective degrees of freedom ${\mathcal N} \sim 100$ ).
For the non-relativistic right-handed neutrino, the Fermi-Dirac equilibrium density $\bar n^{eq}_N$ is well approximated by the Maxwell distribution, yielding in the presence of the background $B_0$: 
\be
\bar n_{N}^{eq}= g_N \, e^{-m_N/T}\,\left(\frac{m_N \, T}{2\pi }\right)^{\frac{3}{2}}+\mathcal{O}(B_0^2/m_N^2)~,
\ee
where $g_N = 2$ is the effective number of degrees of freedom of the right-handed neutrino, and we assume that $B_0/m_N \ll 1$, an assumption that proves to be self consistent.  The lepton asymmetry $\frac{\Delta L^{TOT}}{s}$ has not been measured directly, hence it can - depending on the theory - be different from the baryon asymmetry. However in theories with sphaleron transitions that preserve Baryon-minus-Lepton (B-L) number, such as minimal extensions of the SM with right-handed neutrinos, as the ones we are interested in \cite{decesare} and here, $\Delta L^{TOT}/s$ is expected to be of the same order of magnitude as the baryon asymmetry (\ref{baryon asymmetry}), 
\be \label{baryon asymmetry}
Y_{\Delta B} = \frac{n_{B}-n_{\bar{B}}}{s} = (8.4 - 8.9)\times10^{-11}~, \qquad  T > 1~{\rm GeV}~,
 \ee
where $n_{B}$ ($n_{\bar{B}}$) is the number density of baryons (antibaryons) in the universe, provided it is communicated to the baryon sector by Baryon and Lepton number violating but Baryon-minus-Lepton (B-L) conserving sphaleron processes in the SM sector. An order of magnitude estimate of the ratio $\frac{B_0}{m}$ can be found making use of the approximation $T_D\simeq m_N$ and retaining only first order terms in $\frac{B_0}{m} \ll 1$.
Equating the expression for the lepton asymmetry with the phenomenological value (\ref{baryon asymmetry}), and expanding (\ref{dl}) to first order in $B_0/m_N$, we obtain (for $g_N = 2$)
\be\label{dclept}
\frac{\Delta L^{TOT}}{s} \simeq  \frac{g_N}{ 7 \,e\, (2\pi)^{3/2} }\frac{B_0}{m_N} \simeq 0.007 \, \frac{B_0}{m_N} \simeq 
8 \times 10^{-11}, \qquad T \simeq T_D \simeq m_N~,
\ee
which implies ~\cite{decesare}
\be\label{largeB}
\frac{B_0}{m_N} = \mathcal{O}(10^{-8}).
\ee
The small value of this ratio also allows us to justify \emph{a posteriori} the neglect of higher powers of $B_0$ in the formulae above. For the case where $y = {\mathcal O}(10^{-5})$ and from the lower bound for $m_N$ of 100 TeV found in (\ref{lowerb}), we get an approximation for the smallest possible magnitude of the background field required in order for this mechanism to be effective: $B_0\simeq 1 \; \mbox{MeV}$.  
If other mechanisms contributed to the lepton asymmetry in the universe, or the Yukawa couplings assume smaller values, the minimum value of $B_0$ would be smaller than the one given here. Baryogenesis is then assumed to proceed via B-L conserving processes in the SM sector of the model.

In order to get a  physically correct and more accurate estimate of the induced lepton asymmetry,  the relevant Boltzmann equation needs to be studied in detail, since the heavy right-handed neutrinos are not at rest, but characterised by the Maxwell-Boltzmann momentum distribution in the early universe. This requires a good approximation for the thermally averaged decay rates (\ref{2channels}) of all the relevant processes and will be the subject of the current article. As the Boltzmann equations associated with the leptogenesis scenario advocated here and in \cite{decesare} involve appropriately averaged thermal rates of the decays (\ref{2channels}), we develop in Appendix \ref{sec:App3} the relevant formalism (for $B_0/m_N \ll 1$); the formalism will be used in the next section \ref{sec:boltz} to set up the pertinent system of Boltzmann equations. We shall often borrow methods  and techniques 
from the standard case of \emph{CPT} conserving RHN-induced leptogenesis,  where the CPTV background $B_0$ is absent, but there is CP violation in the lepton sector~\cite{Fukugita-Yanagida,Luty,strumia}. In the current article we shall closely follow the formalism outlined in \cite{strumia}.

\section{Setting up the Boltzmann Equations for Leptogenesis  in the presence of CPTV Backgrounds \label{sec:boltz}}

 In the presence of the weak background $B_0$ the following Boltzmann equation for the number density $n_r$ of a fermion species $\chi$ of mass $m_\chi$ and helicity $\lambda_r$, has been derived in the Appendix of \cite{decesare}:
 \begin{eqnarray}\label{modbolfinal}
&&\dfrac{{\rm d}}{{\rm d}t}\, n_r  + 3Hn_r - \frac{g}{2\pi^2} \, 2\lambda_r \, H\frac{B_0}{T}\, T^3  \int_0^\infty du \, u \, f (E(B_0=0), u) 
=  \\ \nonumber && \frac{g}{8\pi^3}\int \frac{d^3p}{E(B_0 \ne 0)}C[f] + {\mathcal O}(B_0^2/m_N^2)
\end{eqnarray}
where $g$ denotes the number of degrees of freedom, and $f$ is the Fermi Dirac distribution of a relativistic fermion assuming zero chemical potential:
\be
f(E_r; T) = \frac{1}{e^{E_r/T} + 1}
\ee 
The $B_0$ dependent energy-momentum dispersion relation  (\emph{cf.} Appendix \ref{sec:App2})
\begin{align}\label{disp1}
E^{2}_{r}(\vert\bar{p}\vert) = m^{2} + (B_{0} + \lambda_{r}\vert\bar{p}\vert)^{2}
\end{align}
should be used and an expansion up to and including first order terms in the background $B_0/(m_N,T)$ is performed for our weakly CPTV background.  
The term $C[f]$ denotes the appropriate thermally averaged decay or interaction rates involving the species $\chi$~\cite{Kolb}. 
In practice, it is convenient when calculating the lepton asymmetry, to consider the number densities normalised over the entropy density of the universe (\ref{entropydensity})~\cite{Kolb}:
 \begin{equation}\label{yd}
 Y_r \equiv \frac{n_r}{s}~.
 \end{equation}

 In the problem at hand, we consider a system of Boltzmann equations, associated with the heavy neutrino $N$, as well as the lepton $l^\pm$ abundances. 
  The Boltzmann equation (\ref{modbolfinal}) applies to both a relativistic (massless) neutrino as well as a heavy right-handed neutrino, upon using the appropriate dispersion relation (\ref{disp1}). We shall follow the standard analysis in constructing the relevant equations~\cite{strumia}, with the important difference being that the energy momentum dispersion relation and the interaction rates $C[f]$ term involve now the LV and CPTV background $B_0$. 

In terms of the abundances (\ref{yd}), the Boltzmann equations associated with the interactions (\ref{2channels}) of a RHN with a given helicity $\lambda$ take the form:
\begin{align}\label{boltzmanYB}
zHs\dfrac{dY_{N}^{(\lambda)}}{dz} - \lambda I =&\; -\Big\lbrace\gamma^{eq, (\lambda)}(N \rightarrow l^{-}h^{+})\dfrac{Y_{N}^{(\lambda)}}{Y_{N}^{(\lambda),eq}} - \gamma^{eq, (\lambda)}(l^{-}h^{+} \rightarrow N)\dfrac{Y_{l^{-}}^{(\lambda)}}{Y_{l^{-}}^{(\lambda),eq}}\dfrac{Y_{h^{+}}}{Y_{h^{+}}^{eq}}\\
\nonumber
+&\; \gamma^{eq, (\lambda)}(N \rightarrow l^{+}h^{-})\dfrac{Y_{N}^{(\lambda)}}{Y_{N}^{(\lambda),eq}} - \gamma^{eq, (\lambda)}(l^{+}h^{-} \rightarrow N)\dfrac{Y_{l^{+}}^{(\lambda)}}{Y_{l^{+}}^{(\lambda),eq}}\dfrac{Y_{h^{-}}}{Y_{h^{-}}^{eq}}\Big\rbrace~,
\end{align} 
where $Y_N$ is the heavy neutrino abundance, and the superscript $eq$ denotes thermal equilibrium quantities. The equilibrium abundances $Y_N^{eq}$ are discussed in detail in Appendix \ref{sec:App}; the $\gamma^{eq, (\lambda)}(N \stackrel{\leftarrow} \rightarrow \ell^\pm \, h^\mp)$ denote the appropriate thermally averaged decay rates, discussed in Appendices \ref{sec:App2} and \ref{sec:App3}. We shall use their explicit expressions later on, in order to construct the final form of the Boltzmann equations. 
The term $\lambda \, I$ in (\ref{boltzmanYB})  is a generic notation for  
an appropriate integral stemming  from the terms proportional to the CPTV background $B_0$ and the helicity $\lambda$ on the left-hand-side of (\ref{modbolfinal}). Such terms vanish when we average over helicities, since $\sum_r \lambda_r = 0$.  
The reader should notice that apart from the $\lambda\, I$ term, the rest of the structures in (\ref{boltzmanYB}) are the same as 
in conventional \emph{CPT} invariant but CP violating cases for leptogenesis~\cite{strumia}; but, as already mentioned, the relevant dispersion relations (\ref{disp1}) are modified by the CPTV background $B_0\ne 0$.

From the expressions for the relevant amplitudes in Appendix \ref{sec:App2},  we know that, on account of helicity conservation, for the processes $N \stackrel{\leftarrow} \rightarrow l^{-}h^{+}$ we only have one helicity $\lambda = -1$ and for the processes $N \stackrel{\leftarrow} \rightarrow l^{+}h^{-}$ we only have $\lambda = +1$. Following standard treatments~\cite{strumia}, 
we also take the charged Higgs boson as well as the charged leptons to be roughly in equilibrium; hence we set $Y_{l,h} \simeq Y_{i,h}^{eq}$ for the corresponding abundances in (\ref{boltzmanYB}), and find:
 \begin{align}\label{ynbols}
zHs\dfrac{dY_{N}^{(-)}}{dz} + I \simeq&\; -\Big\lbrace\gamma^{eq, (-)}(N \rightarrow l^{-}h^{+})\dfrac{Y_{N}^{(-)}}{Y_{N}^{(-),eq}} - \gamma^{eq, (-)}(l^{-}h^{+} \rightarrow N)\Big\rbrace\\
\nonumber
zHs\dfrac{dY_{N}^{(+)}}{dz} - I \simeq&\; -\Big\lbrace\gamma^{eq, (+)}(N \rightarrow l^{+}h^{-})\dfrac{Y_{N}^{(+)}}{Y_{N}^{(+),eq}} - \gamma^{eq, (+)}(l^{+}h^{-} \rightarrow N)\Big\rbrace~.
\end{align}

Next we will generate the lepton and anti-lepton Boltzmann equations, which are needed in the calculation of the lepton asymmetry. 
As there is only one forward and reverse process for a lepton $l^-$ with a definite helicity $\lambda=-1$, ,
the corresponding Boltzmann equation obtained from
(\ref{modbolfinal}), 
reads
\begin{align}\label{lminusbol2}
zHs\dfrac{dY_{l^{-}}^{(\lambda)}}{dz} - \lambda I =\; -\Big\lbrace\gamma^{eq, (\lambda)}(l^{-}h^{+} \rightarrow N)\dfrac{Y_{l^{-}}^{(\lambda)}}{Y_{l^{-}}^{(\lambda),eq}}\dfrac{Y_{h^{+}}}{Y_{h^{+}}^{eq}}
-\; \gamma^{eq, (\lambda)}(N \rightarrow l^{-}h^{+})\dfrac{Y_{N}^{(\lambda)}}{Y_{N}^{(\lambda),eq}}\Big\rbrace~.
\end{align}
Again we take the Higgs particle to be in equilibrium $Y_{h^{+}} \simeq Y_{h^{+}}^{eq}$~\cite{strumia}. Moreover, from the relevant discussion in Appendix \ref{sec:App2}, we know that we only have one helicity ($\lambda = -1$) for the processes concerning the leptons $l^-$, which implies that the Boltzmann equation for the lepton becomes
\begin{align}\label{lminusbol}
zHs\dfrac{dY_{l^{-}}^{(-)}}{dz} + I \simeq \; -\Big\lbrace\gamma^{eq, (-)}(l^{-}h^{+} \rightarrow N)\dfrac{Y_{l^{-}}^{(-)}}{Y_{l^{-}}^{(-),eq}}
-\; \gamma^{eq, (-)}(N \rightarrow l^{-}h^{+})\dfrac{Y_{N}^{(-)}}{Y_{N}^{(-),eq}}\Big\rbrace
\end{align}
Applying a similar analysis, but now concentrating on the opposite helcity $\lambda = +1$,
we arrive at the Boltzmann equation for the anti-lepton $l^{+}$:
\begin{align}\label{lplusbol}
zHs\dfrac{dY_{l^{+}}^{(+)}}{dz} - I \simeq \; -\Big\lbrace\gamma^{eq, (+)}(l^{+}h^{-} \rightarrow N)\dfrac{Y_{l^{+}}^{(+)}}{Y_{l^{+}}^{(+),eq}}
-\; \gamma^{eq, (+)}(N \rightarrow l^{+}h^{-})\dfrac{Y_{N}^{(+)}}{Y_{N}^{(+),eq}}\Big\rbrace~.
\end{align}

In the specific leptogenesis scenario of \cite{decesare}, the leading contributions to the lepton asymmetry (as far as the small Yukawa coupling (\ref{yc1}), $y \sim 10^{-5} \ll 1 $,  is concerned) come from the \emph{tree level} decays (\ref{2channels}) and their reverse processes. As already mentioned in the previous section, the additional interactions $l h \rightarrow \bar l \bar h $ and $l \bar l \rightarrow h \bar h$, involving a 
tree-level heavy neutrino exchange, are both of higher order in $y$ and suppressed by the heavy mass $m_N$, hence they will be ignored in our case. (It should be remarked that these latter interactions yield contributions comparable to the one loop order graph of fig.~\ref{fig:decays} and hence play an important r\^ole in \emph{CPT} invariant, conventional leptogenesis scenarios~\cite{strumia}). 

From now on, we shall concentrate on constructing the system of Boltzmann equations associated with: 

\noindent (i)  the heavy neutrino abundance in units of entropy density (\emph{cf.} (\ref{yd})),
and averaged over helicities $\lambda=\pm1$:
\be\label{ynave}
\bar{Y}_{N} \equiv \; \dfrac{Y_{N}^{(-)} + Y_{N}^{(+)}}{2}
\ee
and 

\noindent (ii) the lepton-asymmetry for the processes (\ref{2channels}), defined in terms of the lepton abundances:
\begin{align}\label{laave}
\mathcal{L} \equiv& \; Y_{l^{-}}^{(-)} - Y_{l^{+}}^{(+)} = 2\Big[\bar{Y}_{l^{-}} - \bar{Y}_{l^{+}}\Big]~, 
\\ \nonumber \bar{Y}_{l} \equiv& \; \dfrac{Y_{l}^{(-)} + Y_{l}^{(+)}}{2} = \frac{Y_{N}^{(-)} + Y_{N}^{(+)}}{2} = \bar{Y}_{N}~,
\end{align}
where we took into account that the asymmetry is generated between the leptons of helicity $\lambda = -1$ and the anti-leptons of helicity $\lambda = +1$, since these are the only decays for the heavy neutrino (\ref{2channels}), for each of which helicity is conserved. There will be no asymmetry between leptons of helicity $\lambda = +1$ and anti-leptons of helicity $\lambda = -1$ and so $Y_{l^{-}}^{(+)} - Y_{l^{+}}^{(-)} = 0$. Moreover, all of the negative helicity lepton abundance $Y_{l^{-}}$ comes from the decay of the negative helicity heavy neutrino. The same argument for the anti-lepton positive helicity abundance generated by the positive helicity heavy neutrinos. These imply the second of the relations (\ref{laave}). 

The total observable lepton asymmetry, which we want to compute, and compare the result with the estimate (\ref{dclept}), is defined with respect to the corresponding abundances (averaged over helicities) in units of the entropy $s$, as follows:
\begin{framed} 
\begin{align}\label{laave2}
\frac{\Delta L^{TOT}}{s}  \equiv  \dfrac{Y_{l^{-}}^{(-)} - Y_{l^{+}}^{(+)}}{Y_{l^{-}}^{(-)} + Y_{l^{+}}^{(+)}} = \dfrac{\mathcal{L}}{2\bar{Y}_{N}} ~,
\end{align}
\end{framed}\noindent on account of (\ref{laave}). In what follows we proceed with the construction and solution of the Boltzmann equations that correspond to the quantities $\bar Y_N$ and $\mathcal L$. 

To obtain a Boltzmann equation, summed up over helicities, for the averaged RHN abundance $\bar Y_N$ (\ref{laave}) 
from the system (\ref{ynbols}), we sum up these equations, to obtain: 
\begin{align}\label{bolave}
2zHs\dfrac{d\bar{Y}_{N}}{dz} =&\; - \Big\lbrace \gamma^{eq, (-)}(N \rightarrow l^{-}h^{+})\dfrac{Y_{N}^{(-)}}{Y_{N}^{(-),eq}} - \gamma^{eq, (-)}(l^{-}h^{+} \rightarrow N)\\
\nonumber
+&\; \gamma^{eq, (+)}(N \rightarrow l^{+}h^{-})\dfrac{Y_{N}^{(+)}}{Y_{N}^{(+),eq}} - \gamma^{eq, (+)}(l^{+}h^{-} \rightarrow N)\Big\rbrace~.
\end{align}

The asymmetry (\ref{laave2}) will be evaluated at decoupling temperatures 
by solving explicitly the appropriate  system of Boltzmann equations for $\mathcal L$ and $\bar Y_N$ and the result 
will be compared with the estimate  (\ref{dclept}) of \cite{decesare}. In solving the equations we shall approach decoupling by \emph{starting from high temperatures} $T$ and gradually \emph{approaching decoupling} $T \to T_D$ by making use of appropriate approximations (Pad\'e approximants~\cite{pade,padeparticle}), which will allow for analytic expressions for the lepton asymmetry.

In this high-temperature (relativistic) regime, the entropy density of the Universe scales with $T$  as $s \sim 14T^{3}$, whilst the Hubble parameter behaves as~\cite{Kolb}, $H \sim 6T^{2}/M_{pl}$, with $M_{pl}$ the Planck mass. 
Using these relations, we can write 
\begin{align}\label{Tscale}
zHs \sim \; = 84\dfrac{m_{N}^{5}}{M_{pl}\, z^{4}}~, \quad 
HT^{2} \sim \; = 6\dfrac{m_{N}^{4}}{M_{pl}\, z^{4}}~, \quad z \equiv \frac{m_N}{T}~.
\end{align}
The terms $\lambda I$ that appear on the left hand side of the Boltzmann equations (\ref{ynbols}), (\ref{lminusbol}), (\ref{lplusbol}), in the high-temperature regime $T \gg m_\chi$ for a generic fermion of mass $m_\chi$, and degrees of freedom $g_\chi$, can be written as:
\begin{align}\label{Iota}
I_\chi =\; \dfrac{g_{\chi}HB_{0}}{\pi^{2}}\int_{T}^{\infty}d\vert\bar{p}_{\chi}\vert\vert\bar{p}_{\chi}\vert f^{eq}_{\chi}~, \quad 
f^{eq}_{\chi} =\; \dfrac{1}{\exp\Big[\dfrac{E_{\chi}}{T}\Big] + 1} = \exp\Big[-\dfrac{E_{\chi}}{T}\Big]\sum_{n = 0}^{\infty}(-1)^{n}\exp\Big[-n\dfrac{E_{\chi}}{T}\Big]~.
\end{align}
We only have to consider the (massless) lepton case and expand the series upto second order,
\begin{align}
f^{eq}_{l} \simeq&\; \exp\Big[-\dfrac{E_{l}(\vert\bar{p}_{l}\vert)}{T}\Big] + \exp\Big[-2\dfrac{E_{l}(\vert\bar{p}_{l}\vert)}{T}\Big] - \exp\Big[-3\dfrac{E_{l}(\vert\bar{p}_{l}\vert)}{T}\Big],\;\;\;\;\;\;\;\; E_{l}(B_{0} = 0) = \vert\bar{p}_{l}\vert.
\end{align}
The integral $I_{l}$ can therefore be expressed as,
\begin{align}
I_{l} = \dfrac{g_{l}HB_{0}T^{2}}{\pi^{2}}\Big[J_{1} - J_{2} + J_{3}\Big],\;\;\;\; J_{n} = \int_{1}^{\infty}dxxe^{-nx} = \dfrac{n + 1}{n^{2}}e^{-n},
\end{align}
where the integration variable was changed to $\vert\bar{p}_{l}\vert/T = x$. The expression for $I_{l}$ up to second order is given by,
\begin{align}
I_{l} = 1.7842\dfrac{g_{l}HB_{0}T^{2}}{\pi^{2}e}.
\end{align}  
$E_{l}(\vert\bar{p}_{l}\vert)$ is the relativistic energy of the lepton and is taken to be independent of $B_{0}$, since in our analysis we are only considering terms of 
linear order in $B_0 \ll T, m_N$~\cite{decesare}.  All series expansions are taken to second order in the appropriate small 
parameters, for reasons that will become clear below, when we consider the Pad\'e approximated analytic solution for the Boltzmann equations extrapolated to the RHN decoupling 
temperature $T_D \simeq m_N$ (\ref{freeze}), (\ref{hnmass}).

The integral $I_\chi$, in the lepton case, can be approximated by
\begin{align}\label{Ichifinal}
I_{l} \simeq \; 10.7052\, \dfrac{g_{l}\, m_{N}^{4}\, B_{0}}{\pi^{2}\, e\, M_{pl}\, z^{4}}~.
\end{align}
Hence, from (\ref{bolave}), (\ref{lminusbol}), (\ref{lplusbol}), (\ref{Tscale}) and (\ref{Ichifinal}), we observe that the Boltzmann equations for the heavy neutrino abundance and lepton/anti-lepton asymmetry ${\mathcal L}$, averaged over helicities, in the high temperature regime, acquire the form (we reminder the reader that the leptons $l^\pm$ are strictly massless, $m_{l^{\pm}}=0$,  in the high temperature regime, above the electroweak phase transition):
\begin{framed}
\begin{align}\label{bolyn}
168\dfrac{m_{N}^{5}}{M_{pl}\, z^{4}}\dfrac{d\bar{Y}_{N}}{dz} =&\; - \Big\lbrace \gamma^{eq, (-)}(N \rightarrow l^{-}h^{+})\dfrac{Y_{N}^{(-)}}{Y_{N}^{(-),eq}} - \gamma^{eq, (-)}(l^{-}h^{+} \rightarrow N)   \\\nonumber
+&\; \gamma^{eq, (+)}(N \rightarrow l^{+}h^{-})\dfrac{Y_{N}^{(+)}}{Y_{N}^{(+),eq}} - \gamma^{eq, (+)}(l^{+}h^{-} \rightarrow N)\Big\rbrace,\\
\nonumber
\end{align}
and 
\begin{align}\label{deltalbol}
84\dfrac{m_{N}^{5}}{M_{pl}\,z^{4}}\dfrac{d\mathcal{L}}{dz} + 2I_{l} =&\; \gamma^{eq,(-)}(N \rightarrow l^{-}h^{+})\dfrac{Y_{N}^{(-)}}{Y_{N}^{(-), eq}} - \gamma^{eq,(+)}(N \rightarrow l^{+}h^{-})\dfrac{Y_{N}^{(+)}}{Y_{N}^{(+), eq}}  \\ \nonumber
-&\; \Big(\gamma^{eq,(-)}(l^{-}h^{+} \rightarrow N)\dfrac{Y_{l^{-}}^{(-)}}{Y_{l^{-}}^{(-), eq}} - \gamma^{eq,(+)}(l^{+}h^{-} \rightarrow N)\dfrac{Y_{l^{+}}^{(+)}}{Y_{l^{+}}^{(+), eq}}\Big),
\end{align}
with the definitions
\begin{align}\label{defsydl}
\bar{Y}_{N} \equiv&\; \dfrac{Y_{N}^{(-)} + Y_{N}^{(+)}}{2},\;\;\;\;\;\; \mathcal{L} \equiv\; Y_{l^{-}}^{(-)} - Y_{l^{+}}^{(+)},\;\;\;\;\;\; I_{l} \equiv  10.7052\dfrac{g_{l}m_{N}^{4}B_{0}}{\pi^{2}eM_{pl}z^{4}}~.
\end{align}
\end{framed}
We next proceed to solve the above equations which, since they are linear and first-order,  can be in principle exactly solved. However, for the exact solutions to be amenable to analysis, approximations will need to be made; the goal is to find an analytic expression for the lepton asymmetry.

 \subsection{Heavy-Right-Handed-Neutrino abundance Boltzmann equation}

We commence our analysis with the heavy-RHN-Boltzmann equation (\ref{bolyn}). 
The equilibrium populations are calculated in Appendix \ref{sec:App}. The corresponding thermally averaged decay rates  read (see Appendices \ref{sec:App2} and \ref{sec:App3}, and in particular Eq.~(\ref{tadr})):    
\begin{framed}
\begin{align}\label{gamma4}
\gamma^{eq,(-)}(N \rightarrow l^{-}h^{+}) =&\; \gamma^{eq,(-)}(l^{-}h^{+} \rightarrow N) = \; \Lambda f_{1}(z)[1 + \varepsilon_{1}(z)]  \\ \nonumber 
\gamma^{eq,(+)}(N \rightarrow l^{+}h^{-}) =&\;    \gamma^{eq,(+)}(l^{+}h^{-} \rightarrow N)      =\; \Lambda f_{1}(z)[1 - \varepsilon_{1}(z)] 
\end{align}
\end{framed}
where 
\begin{align}\label{lamdadef}
\Lambda =&\; \dfrac{3\vert y\vert^{2}m_{N}^{4}}{16(2\pi)^{3}}\\
\nonumber
f_{1}(z) = z^{-2/3}(0.2553 - 0.1447z^{2} + 0.0957z^{4})~,\;\;&\;\; \varepsilon_{1}(z) = z\dfrac{B_{0}}{m_{N}}\dfrac{0.6062 - 0.3063z^{2}}{0.2553 - 0.1447z^{2} + 0.0957z^{4}}~, \qquad z < 1~. 
\end{align}
The reader should notice the  ``reciprocity'' 
equalities
\begin{align}\label{reverse}
&\gamma^{eq}(l^{-}h^{+} \rightarrow N) = \gamma^{eq}(N \rightarrow l^{-}h^{+}),\;\;\;\;\;\; \gamma^{eq}(l^{+}h^{-} \rightarrow N) = \gamma^{eq}(N \rightarrow l^{+}h^{-})
\end{align}
even in the presence of the CPTV background $B_0 \ne 0$. These are consequences of the equality of the corresponding amplitudes (\ref{ampldiff}) and energy conservation, as explained in 
Appendix \ref{sec:App3}. Also, it is immediately seen from (\ref{gamma4}) that  it is only in the presence of the CPTV background $B_0 \ne 0$ that a lepton asymmetry is  
generated at tree level between the decay channels (\ref{2channels}) (see fig. \ref{fig:decays}), as a consequence of the pertinent differences in (\ref{gamma4}) and (\ref{ampldiff}). 
In this respect, the similarity of the r\^ole of the CPTV $\varepsilon_1$ parameter with the corresponding one, $\varepsilon$,  of conventional leptogenesis~\cite{strumia} should be noticed.The important difference is that, in contrast to our CPTV case, conventional lepton asymmetry occurs at one loop level for the decays  of fig.~\ref{fig:decays} and requires more than one flavour of the RHN.

After substitution of the relevant expression for the thermally-averaged quantities $\gamma^{eq}$, we have the following intermediate results (for details see Appendix \ref{sec:App}), 
\begin{align}\label{int1}
\gamma^{eq, (-)}(l^{-}h^{+} \rightarrow N) + \gamma^{eq, (+)}(l^{+}h^{-} \rightarrow N) = 2\Lambda f_{1}(z),
\end{align}
\begin{align}\label{int2a}
&Y_{N}^{(\lambda), eq} =\; (0.1652)\dfrac{g_{N}}{\pi^{2}e}\Big(1 - 0.176z^{2} + 0.0301z^{4} - 0.9374\lambda\dfrac{B_{0}}{m_{N}}z + 0.2381\lambda \dfrac{B_{0}}{m_{N}}z^{3}\Big),
\end{align}
from which it follows
\begin{align}\label{int2}
&\Big[Y_{N}^{(\lambda), eq}\Big]^{-1} \simeq\; 6.0533\dfrac{\pi^{2}e}{g_{N}}\Big(1 + 0.176z^{2} + 0.0009z^{4} + 0.9374\lambda\dfrac{B_{0}}{m_{N}}z + 0.0919\lambda\dfrac{B_{0}}{m_{N}}z^{3}\Big)\\
\nonumber
&=\; A\Big[g_{1}(z) + \lambda\dfrac{B_{0}}{m_{N}}g_{2}(z)\Big],\\
\nonumber
&A =\; 6.0533\dfrac{\pi^{2}e}{g_{N}}, \;\;\;\; g_{1}(z) = 1 + 0.176z^{2} + 0.0009z^{4}, \;\;\;\; g_{2}(z) = 0.9374z + 0.0919z^{3} ~, \quad z < 1~,
\end{align}  
where to obtain the last expression of (\ref{int2}) we have expanded the function in the round brackets in the definition of $Y_{N}^{(\lambda), eq}$ up to second order in $z < 1$, neglecting terms of order $\mathcal{O}(B_{0}/m_{N})^{2}$. 
The remaining terms in the Boltzmann equation (\ref{bolyn}) become:
\begin{align}\label{eq49}
&\gamma^{eq, (-)}(N \rightarrow l^{-}h^{+})\dfrac{Y_{N}^{(-)}}{Y_{N}^{(-),eq}} + \gamma^{eq, (+)}(N \rightarrow l^{+}h^{-})\dfrac{Y_{N}^{(+)}}{Y_{N}^{(+),eq}} \\
\nonumber
&= \Lambda f_{1}(z)\Big[\dfrac{Y_{N}^{(-)}}{Y_{N}^{(-),eq}} + \dfrac{Y_{N}^{(+)}}{Y_{N}^{(+),eq}} + \varepsilon_{1}(z)\Big(\dfrac{Y_{N}^{(-)}}{Y_{N}^{(-),eq}} - \dfrac{Y_{N}^{(+)}}{Y_{N}^{(+),eq}}\Big)\Big].
\end{align}
We now evaluate the sum and difference of the abundances normalised to their respective equilibrium values,
\begin{align}
\dfrac{Y_{N}^{(-)}}{Y_{N}^{(-),eq}} + \dfrac{Y_{N}^{(+)}}{Y_{N}^{(+),eq}} \simeq&\; A\lbrace g_{1}(z)[Y_{N}^{(-)} + Y_{N}^{(+)}] - \dfrac{B_{0}}{m_{N}}g_{2}(z)[Y_{N}^{(-)} - Y_{N}^{(+)}]\rbrace \simeq 2Ag_{1}(z)\bar{Y}_{N}\\
\nonumber
\dfrac{Y_{N}^{(-)}}{Y_{N}^{(-), eq}} - \dfrac{Y_{N}^{(+)}}{Y_{N}^{(+), eq}} =&\; A[g_{1}(z)(Y_{N}^{(-)} - Y_{N}^{(+)}) - g_{2}(z)\dfrac{B_{0}}{m_{N}}(Y_{N}^{(-)} + Y_{N}^{(+)})]\\
\nonumber
\simeq&\; 2g_{1}(z)z\dfrac{B_{0}}{m_{N}}(0.9374 - 0.2381z^{2}) - 2Ag_{2}(z)\dfrac{B_{0}}{m_{N}}\bar{Y}_{N}\\
\nonumber
Y_{N}^{(-)} - Y_{N}^{(+)} \simeq&\; Y_{N}^{(-) eq} - Y_{N}^{(+) eq} = 2A^{-1}z\dfrac{B_{0}}{m_{N}}(0.9374 - 0.2381z^{2}), \;\;\;\; Y_{N}^{(-)} + Y_{N}^{(+)} = 2\bar{Y}_{N}.
\end{align}
Substituting these expressions in (\ref{eq49}), we obtain
\begin{align}
\gamma^{eq, (-)}(N \rightarrow l^{-}h^{+})\dfrac{Y_{N}^{(-)}}{Y_{N}^{(-),eq}} + \gamma^{eq, (+)}(N \rightarrow l^{+}h^{-})\dfrac{Y_{N}^{(+)}}{Y_{N}^{(+),eq}} \simeq 2A\Lambda f_{1}(z)g_{1}(z)\bar{Y}_{N} + \mathcal{O}\Big(\dfrac{B_{0}}{m_{N}}\Big)^{2},
\end{align} 
where again the term involving the differences of the abundances will be of order $B_{0}^{2}$ since $\varepsilon_{1}(z)$ is already linear in $B_{0}$ and so is neglected. We may write the right-hand-side of the heavy neutrino Boltzmann equation (\ref{bolyn}) as:
\begin{align}
&- \Big\lbrace \gamma^{eq, (-)}(N \rightarrow l^{-}h^{+})\dfrac{Y_{N}^{(-)}}{Y_{N}^{(-),eq}} - \gamma^{eq, (-)}(l^{-}h^{+} \rightarrow N)
+ \gamma^{eq, (+)}(N \rightarrow l^{+}h^{-})\dfrac{Y_{N}^{(+)}}{Y_{N}^{(+),eq}} - \gamma^{eq, (+)}(l^{+}h^{-} \rightarrow N)\Big\rbrace\\
\nonumber
\\
\nonumber
&\simeq - 2A\Lambda f_{1}(z)g_{1}(z)\bar{Y}_{N} + 2\Lambda f_{1}(z)~.
\end{align}
Upon substitution of the relevant expressions, the heavy neutrino Boltzmann equation at high temperatures becomes:
\begin{align}
168\dfrac{m_{N}^{5}}{M_{pl}z^{4}}\dfrac{d\bar{Y}_{N}}{dz} =&\; - 0.2837\dfrac{\vert y\vert^{2}em_{N}^{4}}{g_{N}\pi}z^{-2/3}(0.2553 - 0.0998z^{2} + 0.0704z^{4})\bar{Y}_{N}(z)\\
\nonumber
+&\; \dfrac{3\vert y\vert^{2}m_{N}^{4}}{8(2\pi)^{3}}z^{-2/3}(0.2553 - 0.1447z^{2} + 0.0957z^{4})
\end{align}
which can be finally written as:
\begin{framed}
\begin{align}\label{rhnbe}
&\dfrac{d\bar{Y}_{N}}{dz} + P(z)\bar{Y}_{N} = Q(z)~, \quad z < 1~,\\
\nonumber
\\
\nonumber
&P(z) = a^{2}z^{10/3}\Big(1  - 0.3909z^{2} + 0.2758z^{4}\Big),\;\;\;\;\;\;\;\; a^{2} \equiv \dfrac{0.0724\vert y\vert^{2}eM_{pl}}{168g_{N}\pi m_{N}} \simeq 0.167\\
\nonumber
&Q(z) = b^{2}z^{10/3}\Big(1 - 0.5668z^{2} + 0.3749z^{4}\Big),\;\;\;\;\;\;\;\; b^{2} \equiv \dfrac{0.0957\vert y\vert^{2}M_{pl}}{168(2\pi)^{3}m_{N}} \simeq 0.0056 
\end{align}
\end{framed}
We stress once more that this equation is derived in the high temperature regime in which $m_N < T$.


\subsection{Lepton asymmetry Boltzmann equation}

We proceed now to study the equation for the lepton asymmetry (\ref{deltalbol}) 
at high temperatures. Concentrating on the first two terms on the right hand side, which involve the heavy neutrino abundances, and substituting in the expressions for the
thermally-averaged $\gamma^{eq}$ integrals (\emph{cf.} Appendix \ref{sec:App3}), we obtain after some straightforward manipulations:
\begin{align}
&\gamma^{eq,(-)}(N \rightarrow l^{-}h^{+})\dfrac{Y_{N}^{(-)}}{Y_{N}^{(-), eq}} - \gamma^{eq,(+)}(N \rightarrow l^{+}h^{-})\dfrac{Y_{N}^{(+)}}{Y_{N}^{(+), eq}}\\
\nonumber
&= \Lambda f_{1}(z)\Big[\Big(\dfrac{Y_{N}^{(-)}}{Y_{N}^{(-), eq}} - \dfrac{Y_{N}^{(+)}}{Y_{N}^{(+), eq}}\Big) + \varepsilon_{1}(z)\Big(\dfrac{Y_{N}^{(-)}}{Y_{N}^{(-), eq}} + \dfrac{Y_{N}^{(+)}}{Y_{N}^{(+), eq}}\Big)\Big]\\
\nonumber
&\simeq \Lambda f_{1}(z)\Big[2g_{1}(z)z\dfrac{B_{0}}{m_{N}}(0.9374 - 0.2381z^{2}) - 2Ag_{2}(z)\dfrac{B_{0}}{m_{N}}\bar{Y}_{N} + 2A\varepsilon_{1}(z)g_{1}(z)\bar{Y}_{N} + \mathcal{O}\Big(\dfrac{B_{0}}{m_{N}}\Big)^{3}\Big]~.
\end{align}
where we have substituted in the expressions for the sum and difference of the heavy neutrino abundances from the previous section. The final two terms on the right hand side of the lepton asymmetry Boltzmann equation (\ref{deltalbol}) can be expressed as:
\begin{align}\label{eq56}
&\gamma^{eq,(-)}(l^{-}h^{+} \rightarrow N)\dfrac{Y_{l^{-}}^{(-)}}{Y_{l^{-}}^{(-), eq}} - \gamma^{eq,(+)}(l^{+}h^{-} \rightarrow N)\dfrac{Y_{l^{+}}^{(+)}}{Y_{l^{+}}^{(+), eq}}\\
\nonumber
&= \Lambda f_{1}(z)\Big[\Big(\dfrac{Y_{l^{-}}^{(-)}}{Y_{l^{-}}^{(-), eq}} - \dfrac{Y_{l^{+}}^{(+)}}{Y_{l^{+}}^{(+), eq}}\Big) + \varepsilon_{1}(z)\Big(\dfrac{Y_{l^{-}}^{(-)}}{Y_{l^{-}}^{(-), eq}} + \dfrac{Y_{l^{+}}^{(+)}}{Y_{l^{+}}^{(+), eq}}\Big)\Big]~.
\end{align} 
We next evaluate the sum and difference of the lepton and anti-lepton abundances normalised to their respective equilibrium values, 
that is, the quantities $\dfrac{Y_{l^{-}}^{(-)}}{Y_{l^{-}}^{(-), eq}} \pm \dfrac{Y_{l^{+}}^{(+)}}{Y_{l^{+}}^{(+), eq}}$. Using the explicit expressions for the equilibrium abundances for  leptons and anti-leptons (\emph{cf.} Appendix \ref{sec:App}),
\begin{align}
&Y_{l}^{(\lambda), eq} \simeq (0.1652)\dfrac{g_{l}}{\pi^{2}e}\Big[1 - 0.9374\lambda\dfrac{B_{0}}{m_{N}}z\Big] \Rightarrow \Big[Y_{l}^{(\lambda), eq}\Big]^{-1} \simeq C\Big(1 + 0.9374\lambda\dfrac{B_{0}}{m_{N}}z\Big) + \mathcal{O}\Big(\dfrac{B_{0}}{m_{N}}\Big)^{2},\;\;\;C = (6.0533)\dfrac{\pi^{2}e}{g_{l}}
\end{align}
we obtain
\begin{align}
&\dfrac{Y_{l^{-}}^{(-)}}{Y_{l^{-}}^{(-), eq}} - \dfrac{Y_{l^{+}}^{(+)}}{Y_{l^{+}}^{(+), eq}} = C\Big[\Big(Y_{l^{-}}^{(-)} - Y_{l^{-}}^{(+)}\Big) - 0.9374z\dfrac{B_{0}}{m_{N}}\Big(Y_{l^{-}}^{(-)} + Y_{l^{-}}^{(+)}\Big)\Big] \simeq C\mathcal{L} - 1.8748\dfrac{B_{0}}{m_{N}}z + \mathcal{O}\Big(\dfrac{B_{0}}{m_{N}}\Big)^{2}\\
\nonumber
&\dfrac{Y_{l^{-}}^{(-)}}{Y_{l^{-}}^{(-), eq}} + \dfrac{Y_{l^{+}}^{(+)}}{Y_{l^{+}}^{(+), eq}} \simeq  C\Big[\Big(Y_{l^{-}}^{(-)} + Y_{l^{-}}^{(+)}\Big) - 0.9374z\dfrac{B_{0}}{m_{N}}\Big(Y_{l^{-}}^{(-)} - Y_{l^{-}}^{(+)}\Big)\Big] \simeq 2 - 0.9374C\dfrac{B_{0}}{m_{N}}z\mathcal{L},\\
\nonumber
\\
\nonumber
&Y_{l^{-}}^{(-)} + Y_{l^{+}}^{(+)} \simeq Y_{l^{-}}^{(-), eq} + Y_{l^{+}}^{(+), eq} = 2C^{-1},\;\;\;\;\;\;\;\; \mathcal{L} = Y_{l^{-}}^{(-)} - Y_{l^{+}}^{(+)}.\\
\nonumber 
\end{align}
Then (\ref{eq56}) yields
\begin{align}
&\gamma^{eq,(-)}(l^{-}h^{+} \rightarrow N)\dfrac{Y_{l^{-}}^{(-)}}{Y_{l^{-}}^{(-), eq}} - \gamma^{eq,(+)}(l^{+}h^{-} \rightarrow N)\dfrac{Y_{l^{+}}^{(+)}}{Y_{l^{+}}^{(+), eq}} \simeq \Lambda f_{1}(z)\Big[C\mathcal{L} - 1.8748\dfrac{B_{0}}{m_{N}}z + 2\varepsilon_{1}(z) - \mathcal{O}\Big(\dfrac{B_{0}}{m_{N}}\Big)^{2}\Big]~,
\end{align}
where the reader should recall that $\varepsilon_{1}(z)$ is already linear in $B_{0}/m_{N}$. 

The final form for the lepton-asymmetry Boltzmann equation at high temperatures, then follows:
\begin{framed}
\begin{align}\label{lepbe}
&\dfrac{d\mathcal{L}}{dz} + J(z)\mathcal{L} = H(z)~, \quad z < 1, \\
\nonumber
\\
\nonumber
J(z) =&\; \mu^{2}z^{10/3}\Big(1 - 0.5668z^{2} + 0.3749z^{4}\Big)\\
\nonumber
H(z) =&\; \nu^{2}z^{13/3}\Big(1 - 0.2385z^{2} - 0.3538z^{4}\Big)\bar{Y}_{N}(z) - \sigma^{2}z^{13/3}\Big(1 - 0.1277z^{2} - 1.4067z^{4}\Big) - \delta^{2}\\
\nonumber
\\
\nonumber
\mu^{2} \equiv&\; \dfrac{0.0362\vert y\vert^{2}eM_{pl}}{84g_{l}\pi m_{N}} \simeq 0.227,\;\;\;\;\;\;\;\; \nu^{2} \equiv \dfrac{0.1041\vert y\vert^{2}eM_{pl}B_{0}}{84g_{N}\pi m_{N}^{2}} \simeq 1.3055\dfrac{B_{0}}{m_{N}}\\
\nonumber
\sigma^{2} \equiv&\; \dfrac{0.0479\vert y\vert^{2}M_{pl}B_{0}}{84(2\pi)^{3}m_{N}^{2}} \simeq 0.0056\dfrac{B_{0}}{m_{N}},\;\;\;\;\;\;\;\; \delta^{2} \equiv \dfrac{21.4104}{84}\dfrac{g_{l}B_{0}}{\pi^{2}em_{N}} \simeq 0.038\dfrac{B_{0}}{m_{N}}
\end{align}
\end{framed}
As with the equation for the RHN abundance, the reader should bear in mind that the lepton asymmetry equation above is derived in the high temperature regime $m_N < T$.

\section{Solutions to the System of Boltzmann Equations \label{sec:sol}}

In this section we derive approximate analytic solutions of  the system of Boltzmann equations (\ref{rhnbe}), (\ref{lepbe}), which will allow us to 
compute the lepton asymmetry induced by the CPTV background  in our model. 
So far we have derived equations for the RHN and lepton asymmetry (\emph{cf.} (\ref{rhnbe}) and (\ref{lepbe}) repsectively)
 for high temperatures, $z < 1$. However, we are  eventually interested in solutions of the corresponding Boltzmann equations at the RHN decoupling temperatures (\ref{freeze}), (\ref{hnmass}), where $z \sim 1$~\cite{decesare}. We shall attempt to extrapolate our results above to this case, by 
performing a Taylor expansion of the series solutions to these differential equations. The expansion takes place around an arbitrarily chosen point in the interval $0 < z < 1$, where the solution is valid, taking proper account of the (thermodynamic equilibrium) boundary conditions for the abundances as $z \to 0$ (see Appendix \ref{sec:App}), which fixes the integration constants characterising the solutions. In our analysis below, we take, as a Taylor expansion point, the mid-point of the interval $(0, 1)$, $z=0.5$ . 

To extrapolate the solutions to the regime $z \simeq 1$, 
we shall use a Pad\'e approximation~\cite{pade}.
As well known, 
a Pad\'e expansion can accelerate the convergence of an asymptotic expansion or, for a series, turn a divergence into a convergence. It is widely used for producing in solving approximately complicated problems in several fields of physics, ranging from statistical mechanics to particle physics and quantum field theory~\cite{padeparticle}. Here we present another useful application of the method in cosmology. We outline the general concepts of the Pad\'e approximants method and the specific algorithm used in our computation in this work in Appendix \ref{sec:Pade}.

\subsection{Solution to the Heavy-Neutrino Boltzmann Equation}
The heavy neutrino Boltzmann equation (\ref{rhnbe}) decouples $\bar Y_N$ from $\mathcal L$ so the former can be obtained by solving this equation with an appropriate integrating factor~\cite{4,5}.
We therefore commence our discussion with a sketch of the solution of equation (\ref{rhnbe}).  Calling 
\be\label{abval}
a^2 \equiv \dfrac{0.0724\vert y\vert^{2}eM_{pl}}{168g_{N}\pi m_{N}} \simeq 0.167 \quad {\rm and} \quad  b^2 \equiv \dfrac{0.0957\vert y\vert^{2}M_{pl}}{168(2\pi)^{3}m_{N}} \simeq 0.0056,
\ee
the equation reads
\begin{align}
&\dfrac{d\bar{Y}_{N}}{dz} + P(z)\bar{Y}_{N} = Q(z)~, \quad z < 1~,\\
\nonumber
\\
\nonumber
&P(z) = a^{2}z^{10/3}\Big(1  - 0.3909z^{2} + 0.2758z^{4}\Big)\\
\nonumber
&Q(z) = b^{2}z^{10/3}\Big(1 - 0.5668z^{2} + 0.3749z^{4}\Big)
\end{align}
The integrating factor for this differential equation is given by,
\begin{align}
I_{N}(z) = \exp\Big[\int^{z}dx P(x)\Big] = \exp\Big[a^{2}\Big(\dfrac{3}{13}z^{13/3} - 0.0617z^{19/3} + 0.0331z^{25/3}\Big)\Big].
\end{align}
Multiplying through the differential equation by the integrating factor gives
\begin{align}
\dfrac{d}{dz}\Big[I_{N}(z)\bar{Y}_{N}(z)\Big] = I_{N}(z)Q(z) \;\;\;\; \Rightarrow \;\;\;\; \bar{Y}_{N}(z) = I_{N}^{-1}(z)\Big[c_{1} + \int^{z}dx I_{N}(x)Q(x)\Big]
\end{align}
where $c_{1}$ is the constant of integration and will be determined using the boundary condition (\emph{cf.} (\ref{thermalabund}) in Appendix \ref{sec:App}),
\begin{align}\label{bcYN}
\lim_{z \rightarrow 0}\bar{Y}_{N}(z) \rightarrow \bar{Y}_{N}^{eq} \stackrel{z\to 0} \rightarrow (0.1652)\dfrac{g_{N}}{\pi^{2}e} = \dfrac{b^{2}}{a^{2}} = 5.988b^{2} = 0.0335,
\end{align}
where for heavy right-handed neutrinos $g_N=2$, and we used (\ref{abval}).
In our case $0 < z < 1$, as we are interested in non-trivial populations in the phase where $T > T_D$ (for $T <  T_D$ the populations drop sharply, this is our basic assumption~\cite{decesare}). From the qualitative analysis of \cite{decesare}, reviewed in section \ref{sec:review}, the freezeout temperature $T_D$ is expected to be of order  (\emph{cf.} (\ref{freeze}), (\ref{hnmass})): $T_D \simeq m_N$ so $z_D \simeq 1$. This is why it is important to give formal solutions first, before any expansion. Notice that in arriving at the system of Boltzmann equations for $\bar{Y}_{N}$ and $\mathcal{L}$, we did not make more assumptions on the magnitude of $z$ other than it belongs to the interval $0 < z < 1$.
\begin{framed}
\begin{align}
\bar{Y}_{N}(z) =&\; I_{N}^{-1}(z)\Big[c_{1} + b^{2}\int^{z}dx I_{N}(x)\Big(x^{10/3} - 0.5668x^{16/3} + 0.3749x^{22/3}\Big)\Big], \;\;\;\; 0 < z < 1\\
\nonumber
\\
\nonumber
I_{N}(x) =&\; \exp\Big[a^{2}\Big(\dfrac{3}{13}x^{13/3} - 0.0617x^{19/3} + 0.0331x^{25/3}\Big)\Big]
\end{align}
\end{framed}
We now make some approximations in order to obtain a solution for the heavy neutrino abundance. We can write the integrating factor $I_N(x)$ as
\begin{align}
I_{N}(x) = \exp\Big[\dfrac{3}{13}a^{2}x^{13/3}\Big]S_{n}, \;\;\;\; S_{n} = \sum_{n = 0}^{\infty}\dfrac{(-1)^{n}a^{2n}}{n!}\Big(0.0617 - 0.0331x^{2}\Big)^{n}x^{19n/3}\\
\nonumber
\\
\nonumber
F(z) = b^{2}\int^{z}dx \exp\Big[\dfrac{3}{13}a^{2}x^{13/3}\Big]S_{n}\Big(x^{10/3} - 0.5668x^{16/3} + 0.3749x^{22/3}\Big),
\end{align}
in order to simplify this expression we only take the first two terms in the series $S_{n} \simeq S_{0} + S_{1}$.
\begin{align}
F(z) =&\; b^{2}\int^{z}dx \exp\Big[0.0385x^{13/3}\Big]\Big(1 - 0.0103x^{19/3} + 0.0055x^{25/3}\Big)\Big(x^{10/3} - 0.5668x^{16/3} + 0.3749x^{22/3}\Big)\\
\nonumber
\\
\nonumber
=&\; b^{2}\Big\lbrace \exp\Big[0.0385z^{13/3}\Big]\Big(5.994 - 1.0537z^{2} - 1.136z^{4} + 2.5988z^{6} - 0.9305z^{8} - 93.4633z^{5/3}\\
\nonumber
+&\; 44.6212z^{11/3} - 0.0617z^{19/3} + 0.0677z^{25/3} - 0.042z^{31/3} + 0.0126z^{37/3}\Big) - 399.1316 + 1.0537z^{2}\\
\nonumber
+&\; 1.1365z^{4} + 0.9939z^{6} - 0.7724z^{8} + 93.4657z^{5/3} - 44.6219z^{11/3} + 0.0129z^{19/3} + 0.0212z^{25/3}\Big\rbrace,
\end{align}
where we have expanded again to first order the (upper) incomplete Gamma functions~\cite{8} that arise in this integration,
\begin{align}
\Gamma(s, y) = \Gamma(s) - y^{s}\sum_{k = 0}^{\infty}\dfrac{(-1)^{k}}{k!}\dfrac{y^{k}}{s + k} \simeq&\; \Gamma(s) - s^{-1}y^{s} + [s + 1]^{-1}y^{s + 1},\\
\nonumber
\Gamma\Big(\dfrac{5}{13}, - 0.0385z^{13/3}\Big) \simeq 2.3094 + 0.7429z^{5/3} + 0.0079z^{6},\;\;&\;\; \Gamma\Big(\dfrac{6}{13}, - 0.0385z^{13/3}\Big) \simeq 1.9188 - 0.4819z^{2} - 0.0059z^{19/3}\\
\nonumber
\Gamma\Big(\dfrac{11}{13}, - 0.0385z^{13/3}\Big) \simeq 1.1162 + 0.0751z^{11/3} + 0.0013z^{8},\;\;&\;\;\Gamma\Big(\dfrac{12}{13}, - 0.0385z^{13/3}\Big) \simeq 1.0507 - 0.0536z^{4} - 0.001z^{25/3}.
\end{align}
The boundary condition (\ref{bcYN}) determines the value of the constant of integration: 
$c_{1} = 399.1256b^{2} = 2.2351$. After taking the inverse of the integrating factor (keeping first order terms), 
$$I_{N}^{-1}(z) \simeq \; \exp\Big[-0.0385z^{13/3}\Big]\Big(1 + 0.0103z^{19/3} - 0.0055z^{25/3}\Big),$$
we obtain the expression for the abundance of the heavy neutrino in the interval $0 < z < 1$,
\begin{align}
\bar{Y}_{N}(z < 1) \simeq\; 0.0335 + 0.0001z^{8} - 0.0002z^{19/3} + 0.0001z^{25/3} - 0.0004z^{31/3},
\end{align}
where any exponential factors that remain after multiplying by the inverse of the integrating factor have been expanded to first order. Also any terms of higher order than $z^{32/3}$ have been neglected from the expression due to the restriction $0 < z < 1$ and any terms with factors of order $10^{-5}$ or smaller have also been neglected.

 \subsection{Solution to the Lepton Asymmetry Boltzmann Equation}

 In this subsection, we proceed with the substitution of the previous result onto the Boltzman equation (\ref{lepbe}) and proceed with its solution, which will allow for a determination of the lepton asymmetry.  
Similarly to the previous case, the integrating factor $I_{\mathcal{L}}$ for the lepton asymmetry Boltzmann equation is given by
\begin{align}\label{ifla}
I_{\mathcal{L}}(z) =&\; \exp\Big[\int^{z}dx J(x)\Big] = \exp\Big[\mu^{2}\Big(\dfrac{3}{13}z^{13/3} - 0.0895z^{19/3} + 0.045z^{25/3}\Big)\Big] 
\end{align}
with the lepton asymmetry itself, being expressed as
\begin{align}\label{laif}
\mathcal{L} =&\; I_{\mathcal{L}}^{-1}(z)\Big[c_{2} + \int^{z}dx I_{\mathcal{L}}(x)H(x)\Big],
\end{align} 
where $c_{2}$ is the constant of integration, determined by using the thermal equilibrium boundary condition (\emph{c.f.} Appendix \ref{sec:App}, Eq. (\ref{thermalabund})),  
\begin{align}
\lim_{z \rightarrow 0}\mathcal{L}(z) \rightarrow \mathcal{L}^{eq}(z) \rightarrow 0.
\end{align} 
After substituting in the solution for the $\bar{Y}_{N}(z)$ in the interval $0 < z < 1$ the formal lepton asymmetry solution is given by,
\begin{framed}
\begin{align}
\mathcal{L}(z) =&\; I_{\mathcal{L}}^{-1}(z)\Big[c_{2} + \int^{z}dx I_{\mathcal{L}}(x)\Big(\nu^{2}x^{13/3}\Big(1 - 0.2385x^{2} - 0.3538x^{4}\Big)\bar{Y}_{N}(x) - \sigma^{2}x^{13/3}\Big(1 - 0.1277x^{2} - 1.4067x^{4}\Big) - \delta^{2}\Big)\Big], \nonumber \\
\nonumber
\\
I_{\mathcal{L}}(x) =&\; \exp\Big[\mu^{2}\Big(\dfrac{3}{13}x^{13/3} - 0.0895x^{19/3} + 0.045x^{25/3}\Big)\Big].
\end{align}
\end{framed} 
As in the previous case we make some simplifying approximations to obtain a solution for the lepton asymmetry. The integrating factor is approximated by the expansion of the series up to first order,
\begin{align}
I_{\mathcal{L}}(x) =&\; \exp\Big[\dfrac{3}{13}\mu^{2}x^{13/3}\Big]\sum_{n = 0}^{\infty}\dfrac{(-1)^{n}\mu^{2n}}{n!}\Big(0.0895 - 0.045x^{2}\Big)^{n}x^{19n/3}\\
\nonumber
\simeq&\; \exp\Big[0.0524x^{13/3}\Big]\Big(1 - 0.0047x^{19/3} + 0.0024x^{25/3}\Big).
\end{align}   
Now that an approximate solution for $\bar{Y}_{N}(z)$ is known we may express the coefficient $H(x)$ as,
\begin{align}
H(x) \simeq&\; \dfrac{B_{0}}{m_{N}}\Big(0.0382x^{13/3} - 0.0097x^{19/3} - 0.0076x^{25/3} - 0.0002x^{32/3} - 0.038\Big),
\end{align}
where we have neglected terms of higher powers then $x^{32/3}$. We then have to solve the integral below,
\begin{align}
K(z) =&\; \int^{z}dxI_{\mathcal{L}}(x)H(x)\\
\nonumber
=&\; \dfrac{B_{0}}{m_{N}}\Big\lbrace \exp\Big[0.0524z^{13/3}\Big](0.7467z^{2/3} + 0.1682z - 0.0134z^{3} - 0.0339z^{5} - 0.0009z^{22/3})\\
\nonumber
+&\; 0.8177 - 0.7466z^{2/3} - 0.2063z + 0.0134z^{3} - 0.0052z^{5} - 0.002z^{16/3} + 0.0003z^{22/3}\Big\rbrace,
\end{align}
the (upper) incomplete Gamma functions that appear in the above integration have been evaluated to first order,
\begin{align}
\Gamma\Big(\dfrac{2}{13}, - 0.0524z^{13/3}\Big) \simeq 6.0566 - 4.1294z^{2/3} - 0.0289z^{5},\;\;&\;\; \Gamma\Big(\dfrac{3}{13}, - 0.0524z^{13/3}\Big) \simeq 3.9458 + 2.1942z + 0.0216z^{16/3}\\
\nonumber
\Gamma\Big(\dfrac{9}{13}, - 0.0524z^{13/3}\Big) \simeq 1.3104 + 0.1875z^{3} + 0.004z^{22/3}.
\end{align}
To determine the constant of integration $c_{2}$ we use the boundary condition $\mathcal{L}(z \rightarrow 0) \rightarrow \mathcal{L}^{eq}(z \rightarrow 0) = 0$ which yields $c_{2} = -0.8177\dfrac{B_{0}}{m_{N}}$. Now multiplying by the inverse of the integrating factor (to first order) we obtain an expression for the lepton asymmetry in the interval $0 < z < 1$.
\begin{align}\label{asymmTD}
I_{\mathcal{L}}^{-1}(x) \simeq&\; \exp\Big[-0.0524x^{13/3}\Big]\Big(1 + 0.0047x^{19/3} - 0.0024x^{25/3}\Big)\\
\nonumber
\\
\nonumber
\mathcal{L}(z < 1) =&\; \dfrac{B_{0}}{m_{N}}\Big\lbrace0.0001z^{2/3} - 0.0381z + 0.0088z^{16/3} - 0.0015z^{22/3} + 0.0004z^{28/3} + 0.0001z^{29/3}\Big\rbrace,
\end{align}  
similarly we have neglected terms of higher order powers than $z^{32/3}$ and any terms with factors of order $10^{-5}$ or smaller. Now we want to estimate the lepton asymmetry at freeze out where $T_{D} \leq m_{N}$ corresponding to $z \geq 1$.

To this end we Pade expand~\cite{pade} (\emph{cf.} Appendix \ref{sec:Pade}) the expressions for $\mathcal{L}(z < 1)$ and $\bar{Y}_{N}(z < 1)$ around the point $z = 0.5$ in order to make the expressions for the abundances valid beyond the interval $0 < z < 1$. 
We require a \emph{positive asymmetry} $\mathcal L$, as this is the only physically relevant solution for dominance of matter over antimatter, for our fixed sign of the background $B_0 > 0$.
From (\ref{asymmTD}) we observe that 
$$\mathcal{L}(z < 1.44) < 0~,$$
hence we must have $z = z^\star = 1.44$ as a critical value in our approximate treatment below which the lepton asymmetry switches sign. We interpret this as determining the \emph{freezeout} point, 
\begin{equation}\label{freezezD}
z_D^{\rm Pade} = m_N/T_D \sim 1.44,
\ee
after which ($T < T_D$) the asymmetry freezes out to a positive value. For this value we have
\be\label{finalasymml}
\mathcal{L}(z_D^{\rm Pade} = 1.44) = 0.0009\dfrac{B_{0}}{m_{N}},  \quad \bar{Y}_{N}(z_D^{\rm Pade} = 1.44) = 0.0332~, 
\ee
and thus the observable lepton asymmetry (\ref{laave2})  is given by,
\begin{framed} \begin{align}\label{finalasymmlept}
\dfrac{\Delta L^{TOT}}{s} = \dfrac{\mathcal{L}(z_D = 1.44)}{2\bar{Y}_{N}(z_D = 1.44)} \simeq 0.0136\dfrac{B_{0}}{m_{N}}.
\end{align}
\end{framed}
The reader should compare this result with that obtained in \cite{decesare}, see Eq. (\ref{dclept}) above. Our result (\ref{finalasymmlept}) yields a lepton asymmetry proportional to $B_0/m_N$ 
as in (\ref{dclept}), but with a proportionality
coefficient which is $1.94$ times larger. 
The fact that it is larger may be attributed physically to the fact that here we considered the non zero momentum modes of the heavy neutrino in estimating the asymmetry, which were neglected in \cite{decesare}. Nevertheless, we consider this a good agreement between the two results. 
We have shown above that this lepton asymmetry can be generated at the freeze out point $z = 1.44$ (in order for a positive asymmetry) using first order approximations to the formal solutions of the abundances, this still satisfies the condition that freeze out should occur at $T_{D} \leq m_{N}$.  It is important to notice that the order of magnitude estimate for the Yukawa coupling $\vert y\vert \sim 10^{-5}$ in earlier work ~\cite{decesare}, which was used throughout our previous calculations, providing numerical input (\emph{eg.} (\ref{abval})) into the approximate solutions, remains unchanged, and this provides \emph{a posteriori} a self-consistency check of our approximation. The decoupling (\ref{freezezD}) now occurs at $1.44\,T_{D} = m_{N}$ instead of 
the assumed one in \cite{decesare} at $T_{D} \simeq m_{N}$,  but this does not alter the order of magnitude of the Yukawa coupling. 
However, we believe that the fact that the asymmetry turns negative for $z < 1.44$ is an artefact of the approximations used. Full numerical analysis may lead to a freezeout point $z_D \simeq 1$ as in \cite{decesare}. To check on the stability of the freezeout value, we present next an alternative approximate derivation.

\subsection{Series solutions of the Boltzmann equations}

Here we present another method of obtaining the (approximate) solutions to the differential equations, in an attempt to get an idea on the stability of the freezeout point. Starting with the heavy neutrino Boltzmann equation we can Taylor expand the variable coefficients $P(z), Q(z)$ around the point $z = 0.5$ and the solution $\bar{Y}_{N}(z)$, 
\begin{align}
\bar{Y}_{N}^{\prime}(z) +&\; P(z)\bar{Y}_{N}(z) = Q(z),\\
\nonumber
\\
\nonumber
P(z) = \sum_{n = 0}^{\infty}p_{n}(z - 0.5)^{n},\;\;\;&\;\;\; Q(z) = \sum_{n = 0}^{\infty}q_{n}(z - 0.5)^{n}\\
\nonumber
\bar{Y}_{N}(z) = \sum_{n = 0}^{\infty}c_{n}(z - 0.5)^{n},\;\;\;&\;\;\; \bar{Y}_{N}^{\prime}(z) = \sum_{n = 0}^{\infty}(n + 1)c_{n + 1}(z - 0.5)^{n}.  
\end{align}
On substituting these series into the differential equation we obtain
\begin{align}
&\sum_{n = 0}^{\infty}(n + 1)c_{n + 1}(z - 0.5)^{n} + \Big(\sum_{n = 0}^{\infty}p_{n}(z - 0.5)^{n}\Big)\sum_{m = 0}^{\infty}c_{m}(z - 0.5)^{m} = \sum_{n = 0}^{\infty}q_{n}(z - 0.5)^{n}\\
\nonumber
\Rightarrow &\sum_{n = 0}^{\infty}(n + 1)c_{n + 1}(z - 0.5)^{n} + \sum_{n = 0}^{\infty}\Big(\sum_{k = 0}^{n}c_{k}p_{n - k}\Big)(z - 0.5)^{n} - \sum_{n = 0}^{\infty}q_{n}(z - 0.5)^{n} = 0\\
\nonumber
\Rightarrow &\sum_{n = 0}^{\infty}\Big\lbrace(n + 1)c_{n + 1} + \sum_{k = 0}^{n}c_{k}p_{n - k} - q_{n}\Big\rbrace(z - 0.5)^{n} = 0.
\end{align}
We can then see a recurrence relation for the coefficients of the solution for $\bar{Y}_{N}(z)$ in terms of the coefficients of the $P(z)$ and $Q(z)$ series,
\begin{align}\label{recur}
c_{n + 1} =&\; \dfrac{1}{n + 1}\Big\lbrace q_{n} - \sum_{k = 0}^{n}c_{k}p_{n - k}\Big\rbrace.
\end{align}
Using this recurrence relation, the first few coefficients are:
\begin{align}
\nonumber
p_{0} = P(z)\vert_{z = 0.5} = 0.0152,\;\;\;\; p_{1} = P^{\prime}(z)\vert_{z = 0.5} = 0.0974,\;\;&\;\; p_{2} = \dfrac{1}{2}P^{\prime\prime}(z)\vert_{z = 0.5} = 0.2094,\;\;\;\; p_{3} = \dfrac{1}{6}P^{\prime\prime\prime}(z)\vert_{z = 0.5} = 0.1571\\
\nonumber
q_{0} = Q(z)\vert_{z = 0.5} = 0.0005,\;\;\;\; q_{1} = Q^{\prime}(z)\vert_{z = 0.5} = 0.0031,\;\;&\;\;q_{2} = \dfrac{1}{2}Q^{\prime\prime}(z)\vert_{z = 0.5} = 0.0062,\;\;\;\; q_{3} = \dfrac{1}{6}Q^{\prime\prime\prime}(z)\vert_{z = 0.5} = 0.0039\\
\nonumber
\\
c_{1} = 0.0005 - 0.0152c_{0},\;\;\;\; c_{2} = 0.0015 - 0.0972c_{0},\;\;&\;\; c_{3} = 0.002 - 0.0688c_{0},\;\;\;\; c_{4} = 0.0009 - 0.0359c_{0}.
\end{align}
The Taylor expansion around the point $z = 0.5$ of the heavy neutrino abundance is then,
\begin{align}
\bar{Y}_{N}(z \sim 0.5) = c_{0} + c_{1}(z - 0.5) + c_{2}(z - 0.5)^{2} + c_{3}(z - 0.5)^{3} + c_{4}(z - 0.5)^{4}.
\end{align}
We now take the limit $z \rightarrow 0$ in such a way that the boundary condition (\ref{bcYN}) is satisfied, that is, $\bar{Y}_{N}(z \rightarrow 0) \rightarrow \bar{Y}_{N}^{eq}(z \rightarrow 0) = 0.0335$. This places the final constraint in order to obtain the value for the last remaining coefficient $c_{0} = 0.0339$. The final expression for the heavy neutrino abundance around $z = 0.5$ is given by,
\begin{align}
\bar{Y}_{N}(z \sim 0.5) = 0.0335 + 0.0017z - 0.0018z^{2} + 0.0003z^{3} - 0.0003z^{4}.
\end{align}
We proceed with the analogous calculation for the lepton asymmetry Boltzmann equation,
\begin{align}
\mathcal{L}^{\prime}(z) + J(z)\mathcal{L}(z) = H(z).
\end{align}
The recurrence relation is similar to (\ref{recur}) under the change $p_{n} \rightarrow j_{n},\; q_{n} \rightarrow h_{n},\; c_{n} \rightarrow l_{n}$ where $l_{n}$ are the coefficients in the lepton asymmetry Taylor expansion,
\begin{align}
\mathcal{L}(z) = \sum_{n = 0}^{\infty}l_{n}(z - 0.5)^{n},\;\;\;\;\;\;\;\; l_{n + 1} = \dfrac{1}{n + 1}\Big\lbrace h_{n} - \sum_{k = 0}^{n}l_{k}j_{n - k}\Big\rbrace.
\end{align} 
The coefficients for $J(z)$ and $H(z)$ are:
\begin{align}
\nonumber
j_{0} = J(z)\vert_{z = 0.5} = 0.0199,\;\;\;\; j_{1} = J^{\prime}(z)\vert_{z = 0.5} = 0.1239,\;\;&\;\; j_{2} = \dfrac{1}{2}J^{\prime\prime}(z)\vert_{z = 0.5} = 0.2518,\;\;\;\; j_{3} = \dfrac{1}{6}J^{\prime\prime\prime}(z)\vert_{z = 0.5} =0.1579 \\
\nonumber
h_{0} = H(z)\vert_{z = 0.5} = -0.0362\dfrac{B_{0}}{m_{N}},\;\;&\;\; h_{1} = H^{\prime}(z)\vert_{z = 0.5} = 0.0147\dfrac{B_{0}}{m_{N}}\\
\nonumber
h_{2} = \dfrac{1}{2}H^{\prime\prime}(z)\vert_{z = 0.5} = 0.0441\dfrac{B_{0}}{m_{N}},\;\;&\;\; h_{3} = \dfrac{1}{6}H^{\prime\prime\prime}(z)\vert_{z = 0.5} = 0.0491\dfrac{B_{0}}{m_{N}}.
\end{align}
The coefficients $l_{n}$ are given below using the recurrence relation,
\begin{align}
\nonumber
l_{1} = -0.0362\dfrac{B_{0}}{m_{N}} - 0.0199l_{0},\quad  l_{2} = 0.0077\dfrac{B_{0}}{m_{N}} - 0.0618l_{0}, \quad l_{3} = 0.0161\dfrac{B_{0}}{m_{N}} - 0.0827l_{0}, \quad  l_{4} = 0.0142\dfrac{B_{0}}{m_{N}} - 0.0359l_{0},
\end{align}
which implies
\begin{align}
\mathcal{L}(z \sim 0.5) =&\; l_{0} + l_{1}(z - 0.5) + l_{2}(z - 0.5)^{2} + l_{3}(z - 0.5)^{3} + l_{4}(z - 0.5)^{4}.
\end{align}
We use the boundary condition (\emph{cf.} (\ref{thermalabund}) in Appendix \ref{sec:App}) $\mathcal{L}(z \rightarrow 0) \rightarrow \mathcal{L}^{eq}(z \rightarrow 0) = 0$ to find the last coefficient $l_{0} = - 0.0189\dfrac{B_{0}}{m_{N}}$ and the final expression for the lepton asymmetry is given by,
\begin{align}\label{lepseries}
\mathcal{L}(z \sim 0.5) = \dfrac{B_{0}}{m_{N}}\Big(-0.0389z + 0.0047z^{2} - 0.0121z^{3} + 0.0149z^{4}\Big).
\end{align}
We now perform a Pad\'e expansion~\cite{pade} (\emph{cf.} Appendix \ref{sec:Pade}) around the point $z = 0.5$ to be able to use the solutions outside the interval $0 < z < 1$. In order to obtain a positive asymmetry, we observe from  (\ref{lepseries}) that we must have  $z \ge 1.62$, thus in this approximation the critical point appears to be at $z^* =1.62$. This is identified with the freezeout, 
\be\label{series}
z_D^{\rm series} = 1.62, 
\ee
which, upon substitution into the Pad\'e approximant for the lepton asymmetry, yields $\mathcal{L}(z = 1.62) = 0.0005\dfrac{B_{0}}{m_{N}}$, with the corresponding heavy neutrino abundance at this point is $\bar{Y}_{N}(z = 1.62) = 0.0307$. The observable lepton asymmetry (\ref{laave2}) in that case is found to be 
\begin{framed} \begin{align}\label{laseries}
\dfrac{\Delta L^{TOT}}{s} = \dfrac{\mathcal{L}(z_D = 1.62)}{2\bar{Y}_{N}(z_D = 1.62)} \simeq 0.0081\dfrac{B_{0}}{m_{N}}.
\end{align}
\end{framed}
We see that the series solutions yield a similar answer to the method using an integrating factor. The point of decoupling $z_D = 1.62$ still satisfies $T_{D} \leq m_{N} \Rightarrow z \geq 1$ and the order of magnitude estimate for the Yukawa coupling $\vert y\vert \sim 10^{-5}$ is unchanged. 
Comparing with (\ref{dclept}), we see that the result (\ref{laseries}) is in excellent agreement with the lepton asymmetry estimated in \cite{decesare}. 

From either  (\ref{freezezD}) or  (\ref{laseries}), we obtain that phenomenologically relevant leptogenesis in our system, in the sense of (\ref{dclept}), is achieved for $B_0/m_N ={\mathcal O}(10^{-9}-10^{-8})$, which is in the same approximate range as the estimate of \cite{decesare}, but here 
the result includes all the non-zero momentum modes of the heavy neutrino. This implies that for $m_N ={\mathcal O}(100)$~TeV, we must have 
a $B_0$ in the range  $B_0 \sim 0.1-1~{\rm MeV}$ for leptogenesis to lead to the observed baryogenesis via the B-L conserving sphaleron processes. 

Comparing the freezeout points between the two approximate methods (\ref{freezezD}) and (\ref{series}), we observe agreement with only 12.5 \% uncertainty, indicating stability of the freezeout point in the region around one. 
This completes our analysis. Perhaps as we mentioned earlier, a full numerical solution will yield a freezeout point closer to the qualitative value of \cite{decesare}, although we should emphasize that the above approximate analyses have yielded results in this respect that are of the same order of magnitude. This adds confidence to the efficient application of Pad\'e approximant method to our cosmological problem.

\section{Conclusions and Outlook \label{sec:concl}} 

In this work we have completed the analysis presented in an earlier work~\cite{decesare} by computing the lepton asymmetry generated due to the decays of heavy right-handed neutrinos in the presence of a CPTV axial vector background with only temporal components $B_0 \ne 0$ in the early universe through an analytic (but approximate) solution of the corresponding algebraic system of Boltzmann equations. In \cite{decesare} we only presented a heuristic estimate of the generated asymmetry. The current solution of the Boltzmann equations that describe the leptogenesis in the model has been obtained through an appropriate Pad\'e approximation around the point $z=m_N/T = 0.5$, which allowed the representation of the lepton asymmetry as a power series to be evaluated outside the interval $0 < z < 1$ at the point $z = 1.44$ to generate the positive lepton asymmetry. 

The obtained result for the asymmetry is in \emph{qualitative} agreement with the estimate of \cite{decesare}, in that it is proportional to the small quantity $B_0/m_N  \ll 1$. However  the proportionality coefficient in the case the solutions are evaluated using an integrating factor is found to be $1.94$ times larger than in the case of \cite{decesare}. On the other hand, in case one uses a series solution to the Boltzmann equations, the proportionality coefficient is in excellent agreement with the case of \cite{decesare}. This implies that in our numerical treatment 
the lepton asymmetry can be estimated to be  
\begin{framed}\be\label{tla}
\dfrac{\Delta L^{TOT}}{s} \simeq  (0.008 - 0.014) \dfrac{B_{0}}{m_{N}}, \qquad {\rm at~ freezeout~temperature} \quad T=T_D : \quad m_N/T_D  \simeq (1.44-1.62).
\ee
\end{framed}
This implies that phenomenologically acceptable values of the lepton asymmetry of ${\mathcal O}(8 \times 10^{-11})$ occur for values of 
\begin{framed}\be\label{b0final}
\frac{B_0}{m_N} \sim  10^{-9} - 10^{-8}, \qquad {\rm at~ freezeout~temperature} \quad T=T_D : \quad m_N/T_D  \simeq (1.44-1.62),
\ee
\end{framed}
\noindent in agreement with the estimate (\ref{largeB}) of \cite{decesare}. In our analysis we assumed self-consistently Yukawa couplings in the Higgs portal  term (\ref{yuk}), that couples the right-handed neutrino to the Standard Model sector of the model, of order $|y| \sim 10^{-5}$. This prompted us to ignore higher order terms of order $ \vert y \vert^4 \sim 10^{-20} \ll B_0/m_N$, which 
\emph{a posteriori} was proved to be a self-consistent result, due to the smallness of the $B_0/m_N$ (\ref{b0final}), required for the observed baryon asymmetry today (upon the assumption of the communication of the lepton asymmetry to the baryon sector of the model via B-L conserving sphaleron processes). 

Although our analysis has been generic in not specifying the microscopic origin of the CPTV background, nonetheless some microscopic scenarios originating from string theory have been presented in \cite{decesare}, according to which the background is identified with the dual of the Kalb-Ramond antisymmetric tensor field strength, $\epsilon_{\mu\nu\rho\sigma} \, H^{\nu\rho\sigma}$, which in a four-dimensional space time is equivalent to the derivative of a pseudoscalar field $b(x)$ (Kalb-Ramond axion), $\partial_\mu b$. Nevertheless such an identification is not binding. However, if it is made, then the pressing question concerns the microscopic mechanism, within the context of realistic brane/string models, which underlies the transition from a relatively strong constant (in the Robertson-Walker frame) $B_0 \ne 0$ CPTV background in the early eras of the string Universe, necessary for leptogenesis, to a very weak background today, compatible with the very stringent limits of  CPT Violation in the current era~\cite{smebounds}. Some conjectures to this end have been presented in \cite{decesare} but detailed microscopic mechanisms, compatible with the rest of the asrtroparticle
phenomenology of the models, including the open issue of the smallness of the (observed) cosmological constant (or dark energy) today, are still lacking and constitute the subject of future investigations. 

Nevertheless,  we believe that the scenario for baryogenesis through leptogenesis presented initially in \cite{decesare} and completed here, is an attractive, relatively simple one, which deserves further investigations, within the context of appropriate microscopic models (not necessarily within the framework of string/brane theory). We hope to come back to such studies in the near future. Another important aspect of our current work is the demonstration of the efficiency of the Pad\'e approximant method~\cite{pade} in solving Boltzmann equations, thus adding yet another successful example of this method, this time of relevance to cosmology,

\section*{Acknowledgements}

NEM wishes to thank the University of Valencia and IFIC for a Distinguished Visiting Professorship, during which the current work has been completed. 
The work of TB is supported by an STFC (UK) research studentship and that of NEM and SS is supported in part  by STFC (UK) under the research grant ST/P000258/1. 
 
 \section{Appendices}

In the following Appendices we discuss in detail several technical aspects of our work, which have been used in various parts of the main text.  
 
\subsection{Notation and Conventions}
Throughout this work we use the following conventions. Our metric signature convention is:
\[
g_{\mu\nu} =
\begin{pmatrix}
+1&0&0&0\\
0&-1&0&0\\
0&0&-1&0\\
0&0&0&-1
\end{pmatrix}
\]
which implies 
\[
x^{\mu} =
\begin{pmatrix}
x^{0}\\
\bar{x}
\end{pmatrix}
,\;\;
x_{\mu} = g_{\mu\nu}x^{\nu} =
\begin{pmatrix}
x^{0}\\
-\bar{x}
\end{pmatrix}
\]
\begin{align}
px = p_{\mu}x^{\mu} =&\; g_{\mu\nu}p^{\nu}x^{\mu} = Et - \bar{p}\cdot\bar{x} 
\end{align}
The Dirac $\gamma$ matrices have the properties (we use the symbol $\imath$ to denote the imaginary unit)
\begin{align}
\lbrace\gamma^{\mu}, \gamma^{\nu}\rbrace =&\; \gamma^{\mu}\gamma^{\nu} + \gamma^{\nu}\gamma^{\mu} = 2g^{\mu\nu}\mathbb{1}~, 
 (\gamma^{0})^{2} = \mathbb{1},\;\;\;\; (\gamma^{\jmath})^{2} = -\mathbb{1} \nonumber \\
\nonumber \gamma^{5} =&\; \imath\gamma^{0}\gamma^{1}\gamma^{2}\gamma^{3}~, \quad 
 \lbrace\gamma^{\mu}, \gamma^{5}\rbrace =\; 0~, \quad 
(\gamma^{5})^{2} =\; \mathbb{1}~, \quad 
\gamma^{5\dagger} = \; \gamma^{5}
\end{align}
The chiral representation for the Dirac matrices will be used throughout:
\[
\gamma^{\mu} =
\begin{pmatrix}
0&\sigma^{\mu}\\
\bar{\sigma}^{\mu}&0
\end{pmatrix}
\]
with the $2 \times 2$ Pauli matrices 
\[
\sigma^{\mu} \equiv 
\begin{pmatrix}
\mathbb{1}\\
\sigma^{\jmath}
\end{pmatrix}
,\;\;
\bar{\sigma}^{\mu} \equiv 
\begin{pmatrix}
\mathbb{1}\\
-\sigma^{\jmath}
\end{pmatrix}
\]
\[
\sigma^{1} = 
\begin{pmatrix}
0&1\\
1&0
\end{pmatrix}
,\;\;
\sigma^{2} = 
\begin{pmatrix}
0&-\imath\\
\imath&0
\end{pmatrix}
,\;\;
\sigma^{3} = 
\begin{pmatrix}
1&0\\
0&-1
\end{pmatrix}
\]
and 
\[
\gamma^{5}  =
\begin{pmatrix}
-\mathbb{1}&0\\
0&\mathbb{1}
\end{pmatrix}
\]

\subsection{Decay Amplitudes  \label{sec:App2}} 
 
In this Appendix we work out the amplitudes for the decay channels (\ref{2channels}) in an arbitrary frame, where the decaying right handed neutrino $N$ has a four-momentum $p_\mu$, $\mu=0, \dots 3$. This generalises the approximate treatment of \cite{decesare}, where the field $N$ was assumed at rest.

Our starting point is the Lagrangian for (Dirac) spinors in an axial Background $B_\mu$, which is taken to be 
purely along the temporal axis $(B_{\mu} \rightarrow B_{0})$, with $B_0$ a small, positive (by convention),  non zero constant, $0 < B_{0} \ll 1$:
\begin{align}
\mathcal{L} = \bar{\psi}(\imath\gamma^{\mu}\partial_{\mu} - m\mathbb{1})\psi - \bar{\psi}B_{\mu}\gamma^{\mu}\gamma^{5}\psi.
\end{align}
The corresponding (Dirac) equation of motion reads
\begin{align}\label{diraceq}
(\imath\gamma^{\mu}\partial_{\mu} - m\mathbb{1} - B_{0}\gamma^{0}\gamma^{5})\psi(x) = 0.
\end{align}
On assuming plane-wave solutions for the spinor $\psi$, corresponding to positive ($\psi(x) =\; u(p)e^{-\imath px}$) or negative ($\psi(x) =\; v(p)e^{+\imath px}$)
 frequencies, separately, and substituting in (\ref{diraceq}) we easily obtain~\cite{decesare}
 the pertinent polarization spinors $u(p)$ ($v(p)$) for the positive- (negative) frequency solutions, of helicity $\lambda_r =\pm, \, r=1,2$, in the presence of the background $B_0$ are given by~\cite{decesare} 
\be\label{spinors}
u_{r}(p) = \begin{pmatrix}
\sqrt{E_{r}(\vert\bar{p}\vert) - B_{0} - \lambda_{r}\vert\bar{p}\vert}\xi_{r}\\
\sqrt{E_{r}(\vert\bar{p}\vert) + B_{0} + \lambda_{r}\vert\bar{p}\vert}\xi_{r}
\end{pmatrix}~,
\qquad \qquad 
v_{s}(q) = 
\begin{pmatrix}
\sqrt{E_{s}(\vert\bar{p}\vert) + B_{0} + \lambda_{s}\vert\bar{p}\vert}\xi_{s}\\
-\sqrt{E_{s}(\vert\bar{p}\vert) - B_{0} - \lambda_{s}\vert\bar{p}\vert}\xi_{s}
\end{pmatrix}
\ee
with $u$ ($v$) pertaining to  the (anti) particle, respectively; $\xi_r$ are helicity eigenspinors, satisfying 
\begin{align}\label{helicities}
\frac{\sigma^i \, p^i}{|\vec p|} \, \xi_r = \lambda_r \, \xi_r
\end{align}
with the helicites $\lambda_{1} = -1,\;\; \lambda_{2} = +1$, and $\sigma^i, i=1,2,3$ the $2 \times 2$  Pauli matrices. In the experessions (\ref{spinors})
we used 
the normalisation $\mathcal N^\pm = \sqrt{E_{r}(\vert\bar{p}\vert)  \mp (B_{0} + \lambda_{r}\vert\bar{p})\vert}$, with the $(-)$ $((+))$ sign referring to 
$u$ ($v$) spinors, respectively. 
The eigenspinors $\xi_r$ satisfy the orthogonality condition 
\be\label{ortho}\xi_{s}^{\dagger}\xi_{r} = \delta_{sr}~, \quad s=1,2~.
\ee

The energy-momentum dispersion relation for a fermion of mass $m$ in the presence of $B_0 \ne 0$ reads~\cite{decesare}:
\begin{align}
E^{2}_{r}(\vert\bar{p}\vert) = m^{2} + (B_{0} + \lambda_{r}\vert\bar{p}\vert)^{2}
\end{align}
For the Majorana neutrino we have $m=m_N \ne 0$; on the other hand, the leptons $l^\pm$ in the early Universe, at temperatures much higher than the electroweak symmetry breaking, of interest here, are massless ($m=m_l =0$). Thus, the lepton and neutrino energies are explicitly written as:
\begin{align}\label{energies}
E^{\lambda = -1}_{l^{\pm}}(\vert\bar{p}_{l^{\pm}}\vert) =&\; \vert B_{0} - \vert\bar{p}_{l^{\pm}}\vert\vert ~,
\qquad
E^{\lambda = +1}_{l^{\pm}}(\vert\bar{p}_{l^{\pm}}\vert) =  \vert B_{0} + \vert\bar{p}_{l^{\pm}}\vert\vert = \vert\bar{p}_{l^{\pm}}\vert + B_{0}\\
\nonumber
E^{\lambda = -1}_{N}(\vert\bar{p}_{N}\vert) =&\; \sqrt{m_{N}^{2} + (B_{0} - \vert\bar{p}_{N}\vert)^{2}}~, \qquad \qquad 
E^{\lambda = +1}_{N}(\vert\bar{p}_{N}\vert) =  \sqrt{m_{N}^{2} + (B_{0} + \vert\bar{p}_{N}\vert)^{2}}
\end{align}
Working out the amplitude for the decay process $N \rightarrow l^{-}h^{+}$ we obtain 
\begin{align}\label{amp1}
\imath\mathcal{M}(N \rightarrow l^{-}h^{+}) =&\; -\imath y\bar{u}_{s}(p_{l^{-}})P_{R}u_{r}(p_{N}) = -\imath yu^{\dagger}_{s}(p_{l^{-}})\gamma^{0}P_{R}u_{r}(p_{N})\\
\nonumber
\Rightarrow \mathcal{M}(N \rightarrow l^{-}h^{+}) =&\; - y\xi_{s}^{\dagger}\xi_{r}\sqrt{E_{l^{-},s}(\vert\bar{p}_{l^{-}}\vert) - B_{0} - \lambda_{s}\vert \bar{p}_{l^{-}}\vert}\sqrt{E_{N,r}(\vert\bar{p}_{N}\vert) + B_{0} + \lambda_{r}\vert\bar{p}_{N}\vert}~,
\end{align}
where the outgoing lepton spinor is $\bar{u}_{l^{-},s}(p_{l^{-}})$, and the incoming heavy neutrino spinor $u_{N,r}(p_{N})$;
the notation $E_{\chi, r} (= E^{\lambda_r}_\chi )$ indicates the energy of a spinor $\chi$ with helicity $\lambda_r$.  
$P_R = \frac{1}{2} (1 + \gamma^5)$,  $y$ is the Yukawa coupling (\ref{yc1}) and the orthogonality condition (\ref{ortho}) 
forces the helicities of the incoming and outgoing particles to be the same (\emph{helicity conservation}). After squaring the amplitude (\ref{amp1}) and averaging over initial spins $(S = 1/2)$ we obtain for a given helicity $\lambda$,
\begin{align}\label{a1}
\vert\mathcal{M}\vert^{2}(N \rightarrow l^{-}h^{+}, \lambda) =&\; \dfrac{\vert y\vert^{2}}{2}(E^{\lambda}_{l^{-}} - B_{0} - \lambda\vert\bar{p}_{l^{-}}\vert)(E_{N}^{\lambda} + B_{0} + \lambda\vert\bar{p}_{N}\vert)
\end{align}
We are now going to consider (\ref{a1}) for the two different helicities $\lambda = \pm 1$. The terms within the first bracket of the above expression take the form
\begin{align}
E^{\lambda}_{l^{-}} - B_{0} - \lambda\vert\bar{p}_{l^{-}}\vert = \vert B_{0} + \lambda\vert\bar{p}_{l^{-}}\vert\vert - B_{0} - \lambda\vert\bar{p}_{l^{-}}\vert,
\end{align}
where we have substituted in the lepton energy $E^{\lambda}_{l^{-}} = \vert B_{0} + \lambda\vert\bar{p}_{l^{-}}\vert\vert$. If we take $\lambda = +1$, then the above expression is zero (provided $B_{0} > 0$ which is our initial assumption) 
\begin{align}
E^{\lambda = +1}_{l^{-}} - B_{0} - \vert\bar{p}_{l^{-}}\vert = \vert B_{0} + \vert\bar{p}_{l^{-}}\vert\vert - B_{0} - \vert\bar{p}_{l^{-}}\vert = 0
\end{align}
and so the heavy neutrino  of helicity $\lambda = +1$ can not decay into leptons.

We now consider the case of $\lambda = -1$ for the decay process $N \rightarrow l^{-}h^{+}$.  The terms in the first bracket of the right-hand-side of (\ref{a1}) 
become,
\begin{align}\label{a2}
E^{\lambda = -1}_{l^{-}} - B_{0} + \vert\bar{p}_{l^{-}}\vert = \vert B_{0} - \vert\bar{p}_{l^{-}}\vert\vert - B_{0} + \vert\bar{p}_{l^{-}}\vert~.
\end{align}
We must examine separately the following two cases: (i)  when $\vert\bar{p}_{l^{-}}\vert \leq B_{0}$, the term (\ref{a2}) vanishes, whilst (ii)  when $\vert\bar{p}_{l^{-}}\vert > B_{0}$,
this term becomes $2(\vert\bar{p}_{l^{-}}\vert - B_{0})$.  So for the decay process $N \rightarrow l^{-}h^{+}$, the only way for the amplitude to be non-zero is when $\lambda = -1$ and $\vert\bar{p}_{l^{-}}\vert > B_{0}$. The expression for the amplitude squared  for this process is then given by:
\begin{align}
\vert\mathcal{M}\vert^{2}(N \rightarrow l^{-}h^{+},\; \lambda = -1,\; \vert\bar{p}_{l^{-}}\vert > B_{0}) = \dfrac{\vert y\vert^{2}}{2}\dfrac{m_{N}^{2}}{\vert\bar{p}_{N}\vert}\Big(\vert\bar{p}_{l^{-}}\vert - B_{0}\Big)\Big(1 + \dfrac{B_{0}}{\vert\bar{p}_{N}\vert} - \dfrac{m_{N}^{2}}{4\vert\bar{p}_{N}\vert^{2}}\Big)
\end{align}
where we have substituted in the expression for the relativistic heavy neutrino energy for $\lambda = -1$, expanded up to second order in small quantities, and neglecting terms of order ${\mathcal O}(B_0^2)$  (for our purposes, we assume relativistic regime of temperatures, such that $0 < B_0 \ll m_N \ll p_N \sim T$):
\begin{align}
E_{N}^{\lambda} =&\; \sqrt{m_{N}^{2} + (B_{0} + \lambda\vert\bar{p}_{N}\vert)^{2}}
\simeq\; \vert\bar{p}_{N}\vert + \dfrac{m_{N}^{2}}{2\vert\bar{p}_{N}\vert} - \dfrac{m_{N}^{4}}{8\vert\bar{p}_{N}\vert^{3}} + \lambda\Big(1 - \dfrac{m_{N}^{2}}{2\vert\bar{p}_{N}\vert^{2}}\Big)B_{0}~.
\end{align}
For the reverse process $l^{-}h^{+} \rightarrow N$ we have,
\begin{align}
\imath\mathcal{M}(l^{-}h^{+} \rightarrow N) =&\; -\imath y\overline{P_{R}u}_{N, r}(p_{N})u_{l^{-}, s}(p_{l^{-}})
\end{align} 
where the outgoing heavy neutrino corresponds to $\overline{P_{R}u}_{N, r}(p_{N})$, whilst the incoming lepton to $u_{l^{-}, s}(p_{l^{-}})$. This process yields the same amplitude as for the decay process $N \to l^-\, h^+$, along with the same constraints on the helicity and momentum,
\begin{align}
\vert\mathcal{M}\vert^{2}(l^{-}h^{+} \rightarrow N,\; \lambda = -1,\; \vert\bar{p}_{l^{-}}\vert > B_{0}) = \dfrac{\vert y\vert^{2}}{2}\dfrac{m_{N}^{2}}{\vert\bar{p}_{N}\vert}\Big(\vert\bar{p}_{l^{-}}\vert - B_{0}\Big)\Big(1 + \dfrac{B_{0}}{\vert\bar{p}_{N}\vert} - \dfrac{m_{N}^{2}}{4\vert\bar{p}_{N}\vert^{2}}\Big)
\end{align}
For our purposes in this work, we shall extend the range of the lepton momentum to cover all momenta $\vert\bar{p}_{l^{-}}\vert \in [0, \infty]$. 

For the decay of the heavy neutrino into anti-leptons $N \rightarrow l^{+}h^{-}$ we have the outgoing anti-lepton spinor $v_{l^{+},s}(p_{l^{+}})$ and the incoming heavy neutrino spinor $\bar{v}_{N,r}(p_{N})$ with $N$ being its own anti-particle. The amplitude for this decay is
\begin{align}
\imath\mathcal{M}(N \rightarrow l^{+}h^{-}) =&\; -\imath y\overline{P_{R}v_{N,r}}(p_{N})v_{l^{+},s}(p_{l^{+}}).
\end{align} 
Again we square the amplitude and average over the initial spins of the heavy neutrino, to obtain
\begin{align}
\vert\mathcal{M}\vert^{2}(N \rightarrow l^{+}h^{-}) = \dfrac{\vert y\vert^{2}}{2}(E^{\lambda}_{l^{+}} + B_{0} + \lambda\vert\bar{p}_{l^{+}}\vert)(E_{N}^{\lambda} - B_{0} - \lambda\vert\bar{p}_{N}\vert)
\end{align} 
Consider the energy of the anti-lepton for the possible helicities $E^{\lambda}_{l^{+}} = \vert B_{0} + \lambda\vert\bar{p}_{l^{+}}\vert\vert$. We find that the only two non-zero amplitudes are
\begin{align}
\vert\mathcal{M}\vert^{2}(N \rightarrow l^{+}h^{-}, \lambda = +1) = \vert y\vert^{2}(\vert\bar{p}_{l^{+}}\vert + B_{0})(E_{N}^{\lambda = + 1} - B_{0} - \vert\bar{p}_{N}\vert),\\
\nonumber
\vert\mathcal{M}\vert^{2}(N \rightarrow l^{+}h^{-}, \lambda = -1, \vert\bar{p}_{l^{+}}\vert < B_{0}) = \dfrac{\vert y\vert^{2}}{2}(\vert\bar{p}_{l^{+}}\vert - B_{0})(E_{N}^{\lambda = -1} - B_{0} + \vert\bar{p}_{N}\vert).
\end{align}
We will neglect the contribution from the decay amplitude for negative helicity, as it requires $\vert\bar{p}_{l^{+}}\vert < B_{0}$. Then, for the decay process $N \rightarrow l^{+}h^{-}$ we have
\begin{align}
\vert\mathcal{M}\vert^{2}(N \rightarrow l^{+}h^{-}, \lambda = +1) = \dfrac{\vert y\vert^{2}}{2}\dfrac{m_{N}^{2}}{\vert\bar{p}_{N}\vert}\Big(\vert\bar{p}_{l^{-}}\vert + B_{0}\Big)\Big(1 - \dfrac{B_{0}}{\vert\bar{p}_{N}\vert} - \dfrac{m_{N}^{2}}{4\vert\bar{p}_{N}\vert^{2}}\Big)
\end{align}
where we have substituted in the expansion of $E_{N}^{\lambda = +1}$ up to second order. We see that this decay amplitude differs from the previous process under a change of sign of $B_{0}$. The amplitude for reverse process $l^{+}h^{-} \rightarrow N$ is
\begin{align}
\imath\mathcal{M}(l^{+}h^{-} \rightarrow N) =&\; -\imath y\bar{v}_{l^{+},s}(p_{l^{+}})P_{R}v_{N,r}(p_{N})v_{l^{+},s}(p_{l^{+}}),
\end{align}
and we can readily see that it is the same as that of the forward process. 

The squared amplitudes averaged over initial spins of all the processes are given below:
\begin{framed}
\begin{align}\label{ampldiff}
\vert\mathcal{M}\vert^{2}(N \stackrel{\leftarrow} \rightarrow l^{-}h^{+},\; \lambda = -1) = \dfrac{\vert y\vert^{2}}{2}\dfrac{m_{N}^{2}}{\vert\bar{p}_{N}\vert}\Big(\vert\bar{p}_{l^{-}}\vert - B_{0}\Big)\Big(1 + \dfrac{B_{0}}{\vert\bar{p}_{N}\vert} - \dfrac{m_{N}^{2}}{4\vert\bar{p}_{N}\vert^{2}}\Big)\\
\nonumber
\vert\mathcal{M}\vert^{2}(N \stackrel{\leftarrow} \rightarrow l^{+}h^{-}, \lambda = +1) = \dfrac{\vert y\vert^{2}}{2}\dfrac{m_{N}^{2}}{\vert\bar{p}_{N}\vert}\Big(\vert\bar{p}_{l^{-}}\vert + B_{0}\Big)\Big(1 - \dfrac{B_{0}}{\vert\bar{p}_{N}\vert} - \dfrac{m_{N}^{2}}{4\vert\bar{p}_{N}\vert^{2}}\Big)
\end{align}
\end{framed}where we see that the forward and reverse processes of each decay yield the same result and the difference between the two decay channels into leptons and anti-leptons is a difference in sign of $B_{0}$.

\subsection{Thermal Equilibrium populations \label{sec:App}}

The (thermal) equilibrium population of a particle species is given by~\cite{Kolb}
\begin{align}\label{eqpop}
n^{eq} = g\int\dfrac{d^{3}\bar{p}}{(2\pi)^{3}}f^{eq}
\end{align}
where $f^{eq}$ is the equilibrium distribution function given by Fermi-Dirac or Boson-Einstein statistics.
\begin{align}
f_{l}^{eq} = \dfrac{1}{e^{E_l/T} \pm 1} ~,
\end{align}
with the $+ (-)$ corresponding to fermions (bosons), respectively.

We proceed now to determine  the equilibrium abundances of the heavy right-handed neutrino (RHN) and the leptons. 
In the high-temperature era of the universe that we are considering we have $T > m_{N} \sim T_{D},\; \vert\bar{p}_{N}\vert > m_{N}$ and so the particles behave relativistically. The dispersion relation for the heavy neutrino is given by (\ref{energies}) 
\begin{align}
E_{N}^{(\lambda)}(\vert\bar{p}_{N}\vert) =&\; \sqrt{m_{N}^{2} + (B_{0} + \lambda\vert\bar{p}_{N}\vert)^{2}}\\
\nonumber
=&\; \vert\bar{p}_{N}\vert + \dfrac{m_{N}^{2}}{2\vert\bar{p}_{N}\vert} - \dfrac{m_{N}^{4}}{8\vert\bar{p}_{N}\vert^{3}} - \dfrac{\lambda m_{N}^{2}B_{0}}{2\vert\bar{p}_{N}\vert^{2}} + \lambda B_{0}  
\end{align}
which has been expanded up to second order in small quantities, neglecting terms of ${\mathcal O}(B_0^2)$. From (\ref{eqpop}), then, the equilibrium population is given by
\begin{align}
n_{N}^{(\lambda), eq} =&\; g_{N}\int\dfrac{d^{3}\bar{p}}{(2\pi)^{3}}f_{N}^{eq} = \dfrac{g_{N}}{2\pi^{2}}\int_{T}^{\infty}d\vert\bar{p}_{N}\vert\vert\bar{p}_{N}\vert^{2}f_{N}^{(\lambda), eq},\\
\nonumber
\\
\nonumber
f_{N}^{(\lambda), eq} =&\; \dfrac{1}{\exp\Big[\dfrac{E_{N}^{(\lambda)}}{T}\Big] + 1} = \exp\Big[-\dfrac{E_{N}^{(\lambda)}}{T}\Big]\sum_{n = 0}^{\infty}(-1)^{n}\exp\Big[-n\dfrac{E_{N}^{(\lambda)}}{T}\Big].
\end{align}
Where we expand the series to second order, therefore the equilibrium distribution is approximated by. 
\begin{align}
f_{N}^{(\lambda), eq} \simeq&\; \exp\Big[-\dfrac{E_{N}^{(\lambda)}}{T}\Big] - \exp\Big[-2\dfrac{E_{N}^{(\lambda)}}{T}\Big] + \exp\Big[-3\dfrac{E_{N}^{(\lambda)}}{T}\Big]
\end{align}
the equilibrium population then becomes.
\begin{align}
n_{N}^{(\lambda), eq} =&\; \dfrac{g_{N}}{2\pi^{2}}\int_{T}^{\infty}d\vert\bar{p}_{N}\vert\vert\bar{p}_{N}\vert^{2}\Big(\exp\Big[-\dfrac{E_{N}^{(\lambda)}}{T}\Big] - \exp\Big[-2\dfrac{E_{N}^{(\lambda)}}{T}\Big] + \exp\Big[-3\dfrac{E_{N}^{(\lambda)}}{T}\Big]\Big)
\end{align}
each of the above integrals is of the form,
\begin{align}
&I_{n} = \int_{T}^{\infty}d\vert\bar{p}_{N}\vert\vert\bar{p}_{N}\vert^{2}\exp\Big[-n\dfrac{E_{N}^{(\lambda)}}{T}\Big]\\
\nonumber
&\exp\Big[-n\dfrac{E_{N}^{(\lambda)}}{T}\Big] \simeq \Big[1 - n\Big(\dfrac{m_{N}^{2}}{2\vert\bar{p}_{N}\vert T} - \dfrac{m_{N}^{4}}{8\vert\bar{p}_{N}\vert^{3}T} - \dfrac{\lambda m_{N}^{2}B_{0}}{2\vert\bar{p}_{N}\vert^{2}T}\Big) + \dfrac{n^{2}m_{N}^{4}}{8\vert\bar{p}_{N}\vert^{2}T^{2}}\Big]\exp\Big[- n\dfrac{\lambda B_{0}}{T}\Big]\exp\Big[-n\dfrac{\vert\bar{p}_{N}\vert}{T}\Big]
\end{align}
where $n = 1,2,3$ and we have expanded out the exponential to second order to record all necessary terms,
\begin{align}
I_{n} \simeq&\; T^{3}\exp[-n\dfrac{\lambda B_{0}}{T}]\int_{1}^{\infty}dxx^{2}\Big[1 - n\Big(\dfrac{m_{N}^{2}}{2T^{2}}x^{-1} - \dfrac{m_{N}^{4}}{8T^{4}}x^{-3} - \dfrac{\lambda m_{N}^{2}B_{0}}{2T^{3}}x^{-2}\Big) + \dfrac{n^{2}}{8}\dfrac{m_{N}^{4}}{T^{4}}x^{-2}\Big]\exp[-nx]\\
\nonumber
=&\; \dfrac{T^{3}}{e^{n}}\exp\Big[-n\dfrac{\lambda B_{0}}{T}\Big]\Big\lbrace\dfrac{n^{2} + 2n + 2}{n^{3}} - \dfrac{n + 1}{n}\dfrac{m_{N}^{2}}{2T^{2}} + \dfrac{n}{8}\dfrac{m_{N}^{4}}{T^{4}}\Big[1 + e^{n}\Gamma(0,n)\Big] + \dfrac{\lambda m_{N}^{2}B_{0}}{2T^{3}}\Big\rbrace
\end{align}
above we have changed the integration variable to $\vert\bar{p}_{N}\vert/T = x$ in (117). The result is
\begin{align}
 I_{n} =\; T^{3}e^{-n}\exp\Big[-n\dfrac{\lambda B_{0}}{T}\Big]P_{n}, \;\;\;\; P_{n} = \dfrac{n^{2} + 2n + 2}{n^{3}} - \dfrac{n + 1}{2n}\dfrac{m_{N}^{2}}{T^{2}} + \dfrac{n}{8}\dfrac{m_{N}^{4}}{T^{4}}\Big[1 + e^{n}\Gamma(0,n)\Big] + \dfrac{\lambda m_{N}^{2}B_{0}}{2T^{3}}
 \end{align}
where $\Gamma [s, x] = \int_x^\infty \, t^{s-1}\, e^{-t}\,dt $ is the upper incomplete Gamma function~\cite{8}, with the values
\begin{align}
&\Gamma(0,1) \simeq 0.22, \;\;\;\; \Gamma(0,2) \simeq 0.049, \;\;\;\; \Gamma(0,1) \simeq 0.013
\end{align}
the equilibrium abundance of the RHN (to linear order in $B_{0}$ after expanding the final exponential) is then given 
\begin{align}
n_{N}^{(\lambda), eq} =&\; \dfrac{g_{N}}{2\pi^{2}}\Big(I_{1} - I_{2} + I_{3}\Big) \simeq \dfrac{g_{N}T^{3}}{2\pi^{2}e}\Big[P_{1} - e^{-1}P_{2} + e^{-2}P_{3} - \dfrac{\lambda B_{0}}{T}\Big(P_{1} - 2e^{-1}P_{2} + 3e^{-2}P_{3}\Big)\Big]\\
\nonumber
=&\; \dfrac{5g_{N}T^{3}}{2\pi^{2}e}\Big[0.9251 - 0.1628\dfrac{m_{N}^{2}}{T^{2}} + 0.0278\dfrac{m_{N}^{4}}{T^{4}} - 0.8672\lambda\dfrac{B_{0}}{T} + 0.2203\lambda \dfrac{m_{N}^{2}B_{0}}{T^{3}}\Big]
\end{align}
Next we consider the lepton/antilepton  relativistic abundances. The corresponding dispersion relations (\ref{energies})  are 
(here we do not make a distinction between physical (\emph{i.e.} with positive energies) lepton and anti-lepton excitations as yet, this will be done later)
\begin{align}\label{en1}
E_{l}^{(\lambda)}(\vert\bar{p}_{l}\vert) = \vert B_{0} + \lambda\vert\bar{p}_{l}\vert\vert = \vert\bar{p}_{l}\vert + \lambda B_{0}.
\end{align}
Since we are in the relativistic era and $\vert\bar{p}_{l}\vert \geq T >> B_{0}$, the energy (\ref{en1}) is positive, irrespective of the value of $\lambda$. The corresponding equilibrium populations read:
\begin{align}
n_{l}^{(\lambda), eq} =&\; \dfrac{g_{l}}{2\pi^{2}}\int_{T}^{\infty}d\vert\bar{p}_{l}\vert\vert\bar{p}_{l}\vert^{2}f_{l}^{(\lambda), eq},\\
\nonumber
\\
\nonumber
f_{l}^{(\lambda), eq} =&\; \dfrac{1}{\exp\Big[\dfrac{E_{l}^{(\lambda)}}{T}\Big] + 1} = \exp\Big[-\dfrac{E_{l}^{(\lambda)}}{T}\Big]\sum_{n = 0}^{\infty}(-1)^{n}\exp\Big[-n\dfrac{E_{l}^{(\lambda)}}{T}\Big].
\end{align}
Again we expand the series up to second order. The distribution function is given by, 
\begin{align}
f_{l}^{(\lambda), eq} \simeq&\; \exp\Big[-\dfrac{E_{l}^{(\lambda)}}{T}\Big] - \exp\Big[-2\dfrac{E_{l}^{(\lambda)}}{T}\Big] + \exp\Big[-3\dfrac{E_{l}^{(\lambda)}}{T}\Big]
\end{align}
with the lepton equilibrium abundance (up to second order) being given by
\begin{align}
n_{l}^{(\lambda), eq} =&\; \dfrac{g_{l}}{2\pi^{2}}\Big(J_{1} - J_{2} + J_{3}\Big),\;\;\;\;\;\;\;\; J_{n} = \int_{T}^{\infty}d\vert\bar{p}_{l}\vert\vert\bar{p}_{l}\vert^{2}\exp\Big[-n\dfrac{E_{l}^{(\lambda)}}{T}\Big]
\end{align}
Substituting in the expression for the lepton energy and changing the integration variable $\vert\bar{p}_{l}\vert/T = x$ we obtain,
\begin{align}
J_{n} =&\; T^{3}\exp\Big[-n\dfrac{\lambda B_{0}}{T}\Big]\int_{1}^{\infty}dxx^{2}e^{-nx} =  \dfrac{(n^{2} + 2n + 2)}{n^{3}e^{n}}T^{3}\exp\Big[-n\dfrac{\lambda B_{0}}{T}\Big]
\end{align}
the final expression for the equilibrium lepton abundance is given by, 
\begin{align}
n_{l}^{(\lambda), eq} =&\; \dfrac{g_{l}}{2\pi^{2}}\Big(J_{1} - J_{2} + J_{3}\Big) \simeq \dfrac{5g_{l}T^{3}}{2\pi^{2}e}\Big[0.9251 - 0.8672\lambda\dfrac{B_{0}}{T}\Big].
\end{align}
The difference between the massless lepton and anti-lepton equilibrium abundances will be due to the helicity. 

Of interest to us are the corresponding equilibrium abundances  for RHN ($N$) and leptons ($l$), $Y_{x}^{(\lambda), eq} = n_{x}^{(\lambda), eq}/s, \, x=N, l$ (where $s$ is the entropy density of the Universe that scales with the temperature like $s \sim 14\,T^{3}$), in terms of the quantity $z = m_{N}/T < 1$ (at high $T > m_N)$ , which are: 
\begin{framed}
\begin{align}
Y_{N}^{(\lambda), eq} \simeq&\; (0.1652)\dfrac{g_{N}}{\pi^{2}e}\Big[1 - 0.176z^{2} + 0.0301z^{4} - 0.9374\lambda\dfrac{B_{0}}{m_{N}}z + 0.2381\lambda \dfrac{B_{0}}{m_{N}}z^{3}\Big],\\
\nonumber
Y_{l}^{(\lambda), eq} \simeq&\; (0.1652)\dfrac{g_{l}}{\pi^{2}e}\Big[1 - 0.9374\lambda\dfrac{B_{0}}{m_{N}}z\Big], \\ \nonumber \\
\nonumber z&  \equiv  \frac{m_N}{T}  < 1~, 
\end{align}
\end{framed}  
For our analysis in this work we shall need the averaged over helicities heavy neutrino equilibrium abundance $\bar{Y}_{N}^{eq}$, and the lepton asymmetry equilibrium abundance $\mathcal{L}^{eq}$, which are given by:
\begin{framed}
\begin{align}\label{thermalabund}
\bar{Y}_{N} =&\; \dfrac{1}{2}\Big[Y_{N}^{(-)} + Y_{N}^{(+)}\Big],\;\;\;\;\;\;\;\; \mathcal{L} = Y_{l^{-}}^{(-)} - Y_{l^{+}}^{(+)} \nonumber\\
\nonumber
\\
\bar{Y}_{N}^{eq} =&\; \dfrac{1}{2}\Big[Y_{N}^{(-), eq} + Y_{N}^{(+), eq}\Big] \simeq (0.1652)\dfrac{g_{N}}{\pi^{2}e}\Big(1 - 0.176z^{2} + 0.0301z^{4}\Big)\\
\nonumber
\\
\nonumber
\mathcal{L}^{eq} =&\; Y_{l^{-}}^{(-), eq} - Y_{l^{+}}^{(+), eq} = (0.3097)\dfrac{g_{l}}{\pi^{2}e}\dfrac{B_{0}}{m_{N}}z\\
\nonumber
\\
\nonumber
{\rm with~the~property}& \quad \lim_{z \rightarrow 0}\bar{Y}_{N}^{eq} = (0.1652)\dfrac{g_{N}}{\pi^{2}e},\;\;\;\;\;\;\;\; \lim_{z \rightarrow 0}\mathcal{L}^{eq} = 0
\end{align}
\end{framed}
For heavy right-handed neutrinos, we have $g_N=2$.

\subsection{Thermally averaged Interaction rates \label{sec:App3}}

To calculate the thermal equilibrium density integral for each decay process, which enters the pertinent Boltzmann equation,  we must sum over the different helicities:
\begin{align}
\gamma^{eq}(N \rightarrow l^{\mp}h^{\pm}) = \gamma^{eq,(\lambda = -1)}(N \rightarrow l^{\mp}h^{\pm}) + \gamma^{eq,(\lambda = +1)}(N \rightarrow l^{\mp}h^{\pm})
\end{align}
where, as discussed previously in Appendix \ref{sec:App2}, we will only have the $\lambda = -1$ case for the process $N \rightarrow l^{-}h^{+}$ and the $\lambda = +1$ case for the process $N \rightarrow l^{+}h^{-}$. The interaction integral for the process $N \rightarrow l^{-}h^{+}$ is given by:
\begin{align}\nonumber 
\gamma^{eq,(\lambda = -1)}(N \rightarrow l^{-}h^{+}) =&\; \int\dfrac{d^{3}\bar{p}_{N}}{(2\pi)^{3}2E_{N}}\int\dfrac{d^{3}\bar{p}_{l^{-}}}{(2\pi)^{3}2E_{l^{-}}}\int\dfrac{d^{3}\bar{p}_{h^{+}}}{(2\pi)^{3}2E_{h^{+}}}f_{N}^{eq}(2\pi)^{4}\delta(E_{N} - E_{l^{-}} - E_{h^{+}})\\
\nonumber
\times&\; \delta^{3}(\bar{p}_{N} - \bar{p}_{l^{-}} - \bar{p}_{h^{+}})\vert\mathcal{M}\vert^{2}(N \rightarrow l^{-}h^{+}, \lambda = -1)\\
\nonumber
=&\; \dfrac{1}{8(2\pi)^{5}}\int d^{3}\bar{p}_{N}\int d\vert\bar{p}_{l^{-}}\vert\int d\Omega_{l}\dfrac{f_{N}^{eq}\vert\bar{p}_{l^{-}}\vert^{2}}{E_{N}E_{l^{-}}E_{h^{+}}}
\times \;\delta(f(\vert\bar{p}_{l^{-}}\vert))\vert\mathcal{M}\vert^{2}
\end{align}
with the equilibrium distribution $f_{N}^{eq} = 1/(e^{E_{N}/T} + 1)$. Above, we have integrated over the momentum delta function, to perform explicitly the integration over $d^{3}\bar{p}_{h^{+}}$, which  enforces momentum conservation $\bar{p}_{h^{+}} = \bar{p}_{N} - \bar{p}_{l^{-}}$. The quantity $f(\vert\bar{p}_{l^{-}}\vert)$ is given below
\begin{align}
f(\vert\bar{p}_{l^{-}}\vert) =&\; E_{N}(\vert\bar{p}_{N}\vert) - E_{l^{-}}(\vert\bar{p}_{l^{-}}\vert) - E_{h^{+}}(\vert\bar{p}_{N} - \bar{p}_{l^{-}}\vert)\\
\nonumber
=&\; E_{N} + B_{0} - \vert\bar{p}_{l^{-}}\vert - \Big[\vert\bar{p}_{N}\vert^{2} + \vert\bar{p}_{l^{-}}\vert^{2} - 2\vert\bar{p}_{N}\vert\vert\bar{p}_{l^{-}}\vert\cos(\theta)\Big]^{1/2}\\
\nonumber
\\
\nonumber
f^{\prime}(\vert\bar{p}_{l^{-}}\vert) =&\; - \Big[1 + \dfrac{\vert\bar{p}_{l^{-}}\vert - \vert\bar{p}_{N}\vert\cos(\theta)}{\sqrt{\vert\bar{p}_{l^{-}}\vert^{2} + \vert\bar{p}_{N}\vert^{2} - 2\vert\bar{p}_{N}\vert\vert\bar{p}_{l^{-}}\vert\cos(\theta)}}\Big],
\end{align}
the root of $f(\vert\bar{p}_{l^{-}}\vert)=0$, $\vert\bar{p}_{l^{-}}\vert_{0}$, is:
\begin{align}
\vert\bar{p}_{l^{-}}\vert_{0} = \dfrac{(E_{N} + B_{0})^{2} - \vert\bar{p}_{N}\vert^{2}}{2(E_{N} + B_{0}) - 2\vert\bar{p}_{N}\vert\cos(\theta)} \simeq \dfrac{m_{N}^{2}}{2\vert\bar{p}_{N}\vert[1 - \cos(\theta)]},
\end{align}
where we have only considered the leading term in the expansion of the denominator in the appropriate small quantities. We want to perform the $d\vert\bar{p}_{l^{-}}\vert$ integration in the integral above,
\begin{align}
\int d\vert\bar{p}_{l^{-}}\vert\delta(f(\vert\bar{p}_{l^{-}}\vert)) = \int d\vert\bar{p}_{l^{-}}\vert\dfrac{\delta(\vert\bar{p}_{l^{-}}\vert - \vert\bar{p}_{l^{-}}\vert_{0})}{\vert f^{\prime}(\vert\bar{p}_{l^{-}}\vert_{0})\vert}
\end{align}
which will force $\vert\bar{p}_{l^{-}}\vert \rightarrow \vert\bar{p}_{l^{-}}\vert_{0}$.  The density integral $(\gamma^{eq,(\lambda = -1)}(N \rightarrow l^{-}h^{+}))$ then becomes
\begin{align}\label{forward}
\gamma^{eq,(\lambda = -1)}(N \rightarrow l^{-}h^{+}) =&\; \dfrac{\vert y\vert^{2}m^{2}_{N}}{16(2\pi)^{5}}\int d^{3}\bar{p}_{N}\dfrac{f_{N}^{eq}}{E_{N}\vert\bar{p}_{N}\vert}\Big(1 + \dfrac{B_{0}}{\vert\bar{p}_{N}\vert} - \dfrac{m_{N}^{2}}{4\vert\bar{p}_{N}\vert^{2}}\Big)\\
\nonumber
\times&\;\int d\Omega_{l}\dfrac{\vert\bar{p}_{l^{-}}\vert_{0}^{2}}{\sqrt{\vert\bar{p}_{N}\vert^{2} + \vert\bar{p}_{l^{-}}\vert_{0}^{2} - 2\vert\bar{p}_{N}\vert\vert\bar{p}_{l^{-}}\vert_{0}\cos(\theta)}}\dfrac{1}{\vert f^{\prime}(\vert\bar{p}_{l^{-}}\vert_{0})\vert}.
\end{align}
We now wish to do the angular integration and change the variable $\sin(\theta)d\theta = -d\cos(\theta)$.
\begin{align}\label{I-integral}
I =&\; \int d\Omega_{l}\dfrac{\vert\bar{p}_{l^{-}}\vert_{0}^{2}}{\sqrt{\vert\bar{p}_{N}\vert^{2} + \vert\bar{p}_{l^{-}}\vert_{0}^{2} - 2\vert\bar{p}_{N}\vert\vert\bar{p}_{l^{-}}\vert_{0}\cos(\theta)}}\dfrac{1}{\vert f^{\prime}(\vert\bar{p}_{l^{-}}\vert_{0})\vert}\\
\nonumber
=&\; -2\pi\int_{+1}^{-1}d\cos(\theta)\dfrac{\vert\bar{p}_{l^{-}}\vert^{2}_{0}}{\vert\sqrt{\vert\bar{p}_{N}\vert^{2} + \vert\bar{p}_{l^{-}}\vert_{0}^{2} - 2\vert\bar{p}_{N}\vert\vert\bar{p}_{l^{-}}\vert_{0}\cos(\theta)} + \vert\bar{p}_{l^{-}}\vert_{0} -\vert\bar{p}_{N}\vert\cos(\theta)\vert}\\
\nonumber
\simeq&\; -\dfrac{m_{N}^{4}\pi}{2\vert\bar{p}_{N}\vert^{3}}\int_{+1}^{-1}\dfrac{du}{[1 - u]^{3} + \dfrac{m_{N}^{2}}{2\vert\bar{p}_{N}\vert^{2}}[1 - u]^{2} + \dfrac{m_{N}^{4}}{8\vert\bar{p}_{N}\vert^{4}}}
\end{align}
where we have expanded the square root for $\vert\bar{p}_{N}\vert >\vert\bar{p}_{l^{-}}\vert_{0}$ which is true for most angles and called $\cos(\theta) = u$. Note that the denominator remains always positive. Relabelling $v = 1 - u$, the integral above becomes
\begin{align}
\int_{+1}^{-1}\dfrac{du}{[1 - u]^{3} + \dfrac{m_{N}^{2}}{2\vert\bar{p}_{N}\vert^{2}}[1 - u]^{2} + \dfrac{m_{N}^{4}}{8\vert\bar{p}_{N}\vert^{4}}} = -\int_{0}^{2}\dfrac{dv}{v^{3} + \dfrac{\epsilon^{2}}{2}v^{2} + \dfrac{\epsilon^{4}}{8}}
\end{align}
with $\epsilon = m_{N}/\vert\bar{p}_{N}\vert$. To simplify this integral we will split it up into two regimes where different terms in the denominator are dominant,
\begin{align}
-\int_{0}^{2}\dfrac{dv}{v^{3} + \dfrac{\epsilon^{2}}{2}v^{2} + \dfrac{\epsilon^{4}}{8}} \simeq -\Big[\dfrac{2}{\epsilon^{2}}\int_{0}^{\alpha}\dfrac{dv}{v^{2} + \dfrac{\epsilon^{2}}{4}} + \int_{\alpha}^{2}\dfrac{dv}{v^{3}}\Big]
\end{align}
where $\alpha \simeq \epsilon^{4/3}/2$ denotes the point where $v^{3}$ starts to dominate over the other terms in the denominator. We then have
\begin{align}
-\Big[\dfrac{2}{\epsilon^{2}}\int_{0}^{\alpha}\dfrac{dv}{v^{2} + \dfrac{\epsilon^{2}}{4}} + \int_{\alpha}^{2}\dfrac{dv}{v^{3}}\Big] =&\; \dfrac{1}{8} -2\epsilon^{-8/3} - 4\epsilon^{-3}\tan^{-1}(\epsilon^{1/3})
\simeq\; \dfrac{1}{8} -6\epsilon^{-8/3},
\end{align} 
since $\tan^{-1}(\epsilon^{1/3}) \sim \epsilon^{1/3}$ for $\vert\epsilon^{1/3}\vert << 1$. This implies for the integral in (\ref{I-integral})
\begin{align}
I \simeq -\dfrac{m_{N}\pi}{16}[\epsilon^{3} - 48\epsilon^{1/3}] \simeq 3m_{N}\pi\epsilon^{1/3},\;\;\;\;\;\;\;\; \epsilon = \dfrac{m_{N}}{\vert\bar{p}_{N}\vert}.
\end{align}
Substituting this into the expression for the $\gamma^{eq}$ integral we obtain
\begin{align}
\gamma^{eq,(\lambda = -1)}(N \rightarrow l^{-}h^{+}) =&\; \beta\int_{T}^{\infty} d\vert\bar{p}_{N}\vert\dfrac{f_{N}^{eq}}{E^{(-)}_{N}}\vert\bar{p}_{N}\vert^{2/3}\Big(1 + \dfrac{B_{0}}{\vert\bar{p}_{N}\vert} - \dfrac{m_{N}^{2}}{4\vert\bar{p}_{N}\vert^{2}}\Big),\\
\nonumber
\beta =&\; \dfrac{3\vert y\vert^{2}m^{10/3}_{N}}{16(2\pi)^{3}}.
\end{align}
the expression for the inverse of the heavy neutrino energy is approximated below up to second order, keeping all necessary terms.
\begin{align}
\dfrac{1}{E_{N}^{(-)}} = \dfrac{1}{\vert\bar{p}_{N}\vert\Big[1 + \dfrac{m_{N}^{2}}{2\vert\bar{p}_{N}\vert^{2}} - \dfrac{m_{N}^{4}}{8\vert\bar{p}_{N}\vert^{4}} - \dfrac{B_{0}}{\vert\bar{p}_{N}\vert} + \dfrac{m_{N}^{2}B_{0}}{2\vert\bar{p}_{N}\vert^{3}}\Big]} \simeq \dfrac{1}{\vert\bar{p}_{N}\vert}\Big[1 - \dfrac{m_{N}^{2}}{2\vert\bar{p}_{N}\vert^{2}} + \dfrac{3m_{N}^{4}}{8\vert\bar{p}_{N}\vert^{4}} + \dfrac{B_{0}}{\vert\bar{p}_{N}\vert} - \dfrac{3m_{N}^{2}B_{0}}{2\vert\bar{p}_{N}\vert^{3}}\Big]
\end{align}
substituting this into the integral and multiplying out with the expression in the round brackets we obtain.
\begin{align}
\gamma^{eq,(\lambda = -1)}(N \rightarrow l^{-}h^{+}) =&\; \beta\int_{T}^{\infty} d\vert\bar{p}_{N}\vert\vert\bar{p}_{N}\vert^{-1/3}\Big[1 - \dfrac{3m_{N}^{2}}{4\vert\bar{p}_{N}\vert^{2}} + \dfrac{m_{N}^{4}}{2\vert\bar{p}_{N}\vert^{4}} + \dfrac{2B_{0}}{\vert\bar{p}_{N}\vert} - \dfrac{9m_{N}^{2}B_{0}}{4\vert\bar{p}_{N}\vert^{3}}\Big]f_{N}^{eq}
\end{align}
The equilibrium distribution can be expressed as
\begin{align}
f_{N}^{eq} =&\; \exp\Big[-\dfrac{E_{N}^{(-)}}{T}\Big]\sum_{n = 0}^{\infty}(-1)^{n}\exp\Big[-\dfrac{nE_{N}^{(-)}}{T}\Big]\\
\nonumber
\simeq&\; \exp\Big[-\dfrac{E_{N}^{(-)}}{T}\Big] - \exp\Big[-2\dfrac{E_{N}^{(-)}}{T}\Big] + \exp\Big[-3\dfrac{E_{N}^{(-)}}{T}\Big]
\\
\nonumber
E_{N}^{(-)} =&\; \vert\bar{p}_{N}\vert + \dfrac{m_{N}^{2}}{2\vert\bar{p}_{N}\vert} - \dfrac{m_{N}^{4}}{8\vert\bar{p}_{N}\vert^{3}} - B_{0} + \dfrac{m_{N}^{2}}{2\vert\bar{p}_{N}\vert^{2}}B_{0}
\end{align}
we consider terms upto second order in the exponential series. The gamma integral becomes.
\begin{align}
\gamma^{eq,(\lambda = -1)}(N \rightarrow l^{-}h^{+}) =&\; \beta\int_{T}^{\infty} d\vert\bar{p}_{N}\vert\vert\bar{p}_{N}\vert^{-1/3}\Big[1 - \dfrac{3m_{N}^{2}}{4\vert\bar{p}_{N}\vert^{2}} + \dfrac{m_{N}^{4}}{2\vert\bar{p}_{N}\vert^{4}} + \dfrac{2B_{0}}{\vert\bar{p}_{N}\vert} - \dfrac{9m_{N}^{2}B_{0}}{4\vert\bar{p}_{N}\vert^{3}}\Big]\\
\nonumber
\times&\; \Big(\exp\Big[-\dfrac{E_{N}^{(-)}}{T}\Big] - \exp\Big[-2\dfrac{E_{N}^{(-)}}{T}\Big] + \exp\Big[-3\dfrac{E_{N}^{(-)}}{T}\Big]\Big)
\end{align}
in order to simplify these integrals we notice that there is a general expression,
\begin{align}
&\gamma^{eq,(\lambda = -1)}(N \rightarrow l^{-}h^{+}) = \beta\Big[I_{1} - I_{2} + I_{3}\Big]\\
\nonumber
&I_{n} = \int_{T}^{\infty} d\vert\bar{p}_{N}\vert\vert\bar{p}_{N}\vert^{-1/3}\Big[1 - \dfrac{3m_{N}^{2}}{4\vert\bar{p}_{N}\vert^{2}} + \dfrac{m_{N}^{4}}{2\vert\bar{p}_{N}\vert^{4}} + \dfrac{2B_{0}}{\vert\bar{p}_{N}\vert} - \dfrac{9m_{N}^{2}B_{0}}{4\vert\bar{p}_{N}\vert^{3}}\Big]\exp\Big[-n\dfrac{E_{N}^{(-)}}{T}\Big]\\
\nonumber
\\
\nonumber
&\exp\Big[-n\dfrac{E_{N}^{(-)}}{T}\Big] \simeq \Big[1 - n\Big(\dfrac{m_{N}^{2}}{2\vert\bar{p}_{N}\vert T} - \dfrac{m_{N}^{4}}{8\vert\bar{p}_{N}\vert^{3}T} + \dfrac{m_{N}^{2}B_{0}}{2\vert\bar{p}_{N}\vert^{2}T}\Big) + \dfrac{n^{2}m_{N}^{4}}{8\vert\bar{p}_{N}\vert^{2}T^{2}}\Big]\exp\Big[n\dfrac{B_{0}}{T}\Big]\exp\Big[-n\dfrac{\vert\bar{p}_{N}\vert}{T}\Big]
\end{align}
where $n = 1, 2, 3$ and again expanding the exponential to second order. We thus obtain an expression for the integral $I_{n}$:
\begin{align}
I_{n} =&\; T^{2/3}\exp\Big[n\dfrac{B_{0}}{T}\Big]J_{n}\\
\nonumber
\\
\nonumber
J_{n} =&\; \int_{1}^{\infty}dxx^{-1/3}\Big[1 + \Big(\dfrac{2B_{0}}{T} - \dfrac{nm_{N}^{2}}{2T^{2}}\Big)x^{-1} + \Big(\dfrac{n^{2}m_{N}^{4}}{8T^{4}} - \dfrac{3m_{N}^{2}}{4T^{2}} - \dfrac{3nm_{N}^{2}B_{0}}{2T^{3}}\Big)x^{-2} \nonumber \\
&+ \Big(\dfrac{nm_{N}^{4}}{2T^{4}} - \dfrac{9m_{N}^{2}B_{0}}{4T^{3}}\Big)x^{-3} + \dfrac{m_{N}^{4}}{2T^{4}}x^{-4}\Big]e^{-nx}
\end{align}
where we have employed a change of variable $\vert\bar{p}_{N}\vert/T = x$. Substituting in the different values for $n$, we obtain the solutions for $J_{n}$, after performing the appropriate integrations:
\begin{align}
J_{1} =&\; \dfrac{1}{1120eT^{4}}\Big\lbrace -63m_{N}^{4}\Big[- 6 + e\Gamma\Big(\dfrac{2}{3}, 1\Big)\Big] + 1120T^{3}\Big[6B_{0} + e(-6B_{0} + T)\Gamma\Big(\dfrac{2}{3}, 1\Big)\Big]\\
\nonumber
-&\; 30m_{N}^{2}T\Big[6B_{0} + 14T + e(45B_{0} + 7T)\Gamma\Big(\dfrac{2}{3}, 1\Big)\Big]\Big\rbrace\\
\nonumber
\\
\nonumber
J_{2} =&\; \dfrac{1}{560e^{2}T^{4}}\Big\lbrace m_{N}^{4}\Big[462 -504e^{2}E_{1/3}(2)\Big] + 560T^{3}\Big[6B_{0} + e^{2}(-12B_{0} + T)E_{1/3}(2)\\
\nonumber
-&\; 15m_{N}^{2}T\Big[-114B_{0} + 7T + 4e^{2}(90B_{0} + 7T)E_{1/3}(2)\Big]\Big\rbrace\\
\nonumber
\\
\nonumber
J_{3} =&\; \dfrac{1}{8T^{4}}\Big\lbrace \dfrac{1}{e^{3}}\Big[\dfrac{78}{5}m_{N}^{4} + \dfrac{486}{7}m_{N}^{2}B_{0}T + 48B_{0}T^{3}\Big] - \dfrac{1}{140}\Big[5103m_{N}^{4} + 1120(18B_{0} - T)T^{3} + 270m_{N}^{2}T(135B_{0} + 7T)\Big]E_{1/3}(3)\Big\rbrace\\
\end{align}
with $E_n(x) = \int_1^\infty \, dy \, y^{-n}\, e^{-x\, y}$ is the exponential integral function~\cite{8}, 
and we have the values
\begin{align}
\Gamma\Big(\dfrac{2}{3}, 1\Big) = E_1(1) \simeq 0.304429,\;\;\;\; E_{1/3}(2) \simeq 0.0602489,\;\;\;\; E_{1/3}(3) \simeq 0.015246.
\end{align} 
On substituting $m_{N}/T = z$, we obtain
\begin{align}
J_{1} =&\; 0.3044 - 0.1959z^{2} + 0.107z^{4} + z\dfrac{B_{0}}{m_{N}}\Big(0.3807 - 0.4281z^{2}\Big)\\
\nonumber
J_{2} =&\; 0.0603 - 0.0701z^{2} + 0.0574z^{4} + z\dfrac{B_{0}}{m_{N}}\Big(0.089 - 0.1666z^{2}\Big)\\
\nonumber
J_{3} =&\; 0.0112 - 0.0189z^{2} + 0.0461z^{4} + z\dfrac{B_{0}}{m_{N}}\Big(0.0971 + 0.0676z^{2}\Big).
\end{align}
Therefore to obtain the expression for the $\gamma^{eq}$ integral we recall that,
\begin{align}
&I_{n} = m_{N}^{2/3}z^{-2/3}\exp\Big[n\dfrac{B_{0}}{m_{N}}z\Big]J_{n}\\
\nonumber
&\gamma^{eq,(\lambda = -1)}(N \rightarrow l^{-}h^{+}) = \beta\Big[I_{1} - I_{2} + I_{3}\Big] \simeq \beta m_{N}^{2/3}z^{-2/3}\Big[J_{1} - J_{2} + J_{3} + z\dfrac{B_{0}}{m_{N}}\Big(J_{1} - 2J_{2} + 3J_{3}\Big)\Big].
\end{align}
This implies that the $\gamma^{eq}$ integral (to linear order in $B_{0}$) becomes
\begin{align}
\gamma^{eq,(\lambda = -1)}(N \rightarrow l^{-}h^{+}) \simeq&\; \dfrac{3\vert y\vert^{2}m^{4}_{N}}{16(2\pi)^{3}}z^{-2/3}\Big[0.2553 - 0.1447z^{2} + 0.0957z^{4} + z\dfrac{B_{0}}{m_{N}}\Big(0.6062 - 0.3063z^{2}\Big)\Big]
\end{align}
We next proceed to obtain the expression of the pertinent $\gamma^{eq}$-integral for the reverse process $l^{-}h^{+} \rightarrow N$.
The steps will parallel those of the previous calculation, the only difference is that now one should make the substitution $f_{N}^{eq} \rightarrow f_{l^{-}}^{eq}f_{h^{+}}^{eq}$ in (\ref{forward}).
We thus have 
\begin{align}\label{2gammaintegral}
\gamma^{eq,(\lambda = -1)}(l^{-}h^{+} \rightarrow N) =&\; \dfrac{\vert y\vert^{2}m^{2}_{N}}{16(2\pi)^{5}}\int d^{3}\bar{p}_{N}\dfrac{1}{E_{N}\vert\bar{p}_{N}\vert}\Big(1 + \dfrac{B_{0}}{\vert\bar{p}_{N}\vert} - \dfrac{m_{N}^{2}}{4\vert\bar{p}_{N}\vert^{2}}\Big)\\
\nonumber
\times&\;\int d\Omega_{l}\dfrac{\vert\bar{p}_{l^{-}}\vert_{0}^{2}f_{l^{-}}^{eq}f_{h^{+}}^{eq}}{\sqrt{\vert\bar{p}_{N}\vert^{2} + \vert\bar{p}_{l^{-}}\vert_{0}^{2} - 2\vert\bar{p}_{N}\vert\vert\bar{p}_{l^{-}}\vert_{0}\cos(\theta)}}\dfrac{1}{\vert f^{\prime}(\vert\bar{p}_{l^{-}}\vert_{0})\vert}\\
\nonumber
\\
\nonumber
f_{l^{-}}^{eq} =&\; \dfrac{1}{\exp\Big[\dfrac{E_{l^{-}}^{(-)}}{T}\Big] + 1} = \exp\Big[-\dfrac{E_{l^{-}}^{(-)}}{T}\Big]\sum_{n = 0}^{\infty}(-1)^{n}\exp\Big[-n\dfrac{E_{l^{-}}^{(-)}}{T}\Big],\\
\nonumber
f_{h^{+}}^{eq} =&\; \dfrac{1}{\exp\Big[\dfrac{E_{h^{+}}}{T}\Big] - 1} = \exp\Big[-\dfrac{E_{h^{+}}}{T}\Big]\sum_{n = 0}^{\infty}\exp\Big[-n\dfrac{E_{h^{+}}}{T}\Big],
\end{align}
where, as in the previous case,  in the equilibrium distributions we keep only the first term in the series, which implies
\be\label{product}
f_{l^{-}}^{eq}\, f_{h^{+}}^{eq}  \simeq \exp\Big[-\dfrac{E_{l^{-}}^{(-)} + E_{h^{+}}}{T}\Big].
\ee
On using energy and helicity ($\lambda=-1$)  conservation  in the reaction $l^{-}h^{+} \rightarrow N$, we observe that 
the numerator of the fraction in the exponent in (\ref{product}) can be replaced by the energy of the RHN $E_N^{(\lambda=-1)}$. Then, upon approximating 
(in the high temperature regime) $\exp\Big[-\dfrac{E_N^{(\lambda=-1})}{T}\Big] \simeq  f_{N}^{eq}$, we may write
\be\label{productRHN}
f_{l^{-}}^{eq}\, f_{h^{+}}^{eq} = f_{N}^{eq}~,
\ee
which, upon substitution in (\ref{2gammaintegral}) and comparison with (\ref{forward}), implies the reciprocity (chemical equilibrium) relation for the thermally averaged decay rates, 
\begin{align}\label{reversed}
\gamma^{eq,(\lambda = -1)}(l^{-}h^{+} \rightarrow N) = 
\gamma^{eq,(\lambda = -1)}(N \rightarrow l^{-}h^{+})~, 
\end{align}
in the presence of CPTV background $B_0 \ne 0$. 

The results for the decay and reverse processes $N \stackrel{\leftarrow} \rightarrow l^{+}h^{-}$ will be analogous to those of the previous calculations but with a change in the sign of $B_{0}$, due to the opposite helicity $\lambda=+1$ involved in those processes. The results for all thermally averaged decay rates are summarized below:
\begin{framed}
\begin{align}\label{tadr}
\nonumber
\gamma^{eq,(\lambda = -1)}(N \rightarrow l^{-}h^{+}) =   \gamma^{eq,(\lambda = -1)}(l^{-}h^{+} \rightarrow N)  = &\; \dfrac{3\vert y\vert^{2}m_{N}^{4}}{16(2\pi)^{3}}z^{-2/3}\Big(0.2553 - 0.1447z^{2} + 0.0957z^{4}\Big)\\
\nonumber
\times&\;\Big[1 + z\dfrac{B_{0}}{m_{N}}\dfrac{0.6062 - 0.3063z^{2}}{0.2553 - 0.1447z^{2} + 0.0957z^{4}}\Big]\\
\nonumber
\\
\gamma^{eq,(\lambda = +1)}(N \rightarrow l^{+}h^{-}) =    \gamma^{eq,(\lambda = +1)}(l^{+}h^{-} \rightarrow N) =& \; \dfrac{3\vert y\vert^{2}m_{N}^{4}}{16(2\pi)^{3}}z^{-2/3}\Big(0.2553 - 0.1447z^{2} + 0.0957z^{4}\Big) \nonumber \\
\times&\;\Big[1 - z\dfrac{B_{0}}{m_{N}}\dfrac{0.6062 - 0.3063z^{2}}{0.2553 - 0.1447z^{2} + 0.0957z^{4}}\Big]~.
\end{align}
\end{framed}
Eq.~(\ref{tadr}) implies the generation of a lepton asymmetry between the decay channels (\ref{2channels}) of fig.~\ref{fig:decays} at \emph{tree level} only when $B_0 \ne 0$, due to the difference in the respective decay rates. 

\section{Pad\'e Approximants Method \label{sec:Pade}}

It is often possible to  increase our knowledge of a function $f( z)$ beyond 
the region of convergence of its Taylor series using the method of Pad\'e approximants.  The Pad\'e approximation~\cite{pade} can be considered as follows: given a function $f(z)$ (with a Taylor expansion around $z=0$), and two non-negative integers $m, n \ge 0$, the \emph{Pad\'e approximant} $[m/n]_{f(z)}$ is provided by the function
\be\label{pade}
 \mathcal{P}^{n}_{m}(z)= \frac{\sum_{i=o}^{m} a_i \, z^i }{1 + \sum_{j=1}^n b_k\, z^k} .
\ee 
If the Taylor expansion of $f(z)$ is truncated at power $z^{n+m}$, the resulting polynomial 
${T_{m + n}}\left( z \right)$ can be written as 
\be\label{TaylorfP}
{T_{m + n}}\left( z \right) = \sum\limits_{j = 0}^{m + n} {{c_j}{z^j}} .
\ee
We can Taylor expand  $\mathcal{P}^{n}_{m}(z)$ to order $z^{n+m}$ and equate the expression to  $T_{m + n}\left( z \right)$. (As $n,m\to \infty$ often $\mathcal{P}^{n}_{m}(z)\to f(z)$ even when the Taylor series for $f(z)$ is divergent.) Let us consider a $m\times m$ matrix $\mathcal{A}$ defined by $\mathcal{A}_{ij}=c_{n+i-j}\,(1\leqslant i,j \leqslant m)$. From the matching of the series the $b_{i}$ satisfy the matrix equation
\begin{equation}
\label{ Padecoeff}
\mathcal{A}\left[\begin{array}{c}
b_{1}\\
b_{1}\\
.\\
.\\
.\\
b_{m}
\end{array}\right]=-\left[\begin{array}{c}
c_{n+1}\\
c_{n+2}\\
.\\
.\\
.\\
c_{n+m}
\end{array}\right]
 \end{equation}
The coefficients $a_j, \, b_j$ in (\ref{pade}) are \emph{uniquely} determined, provided we normalise the zeroth order term in the denominator to one.  
The coefficients $a_{i}$ are determined by the set of equations
\begin{equation}
\label{ Padecoeff2}
A_{i}=\sum_{j=0}^{i}c_{i-j}b_{j}.
\end{equation}
A common procedure is to examine the convergence of the sequence $\mathcal{P}^{J}_{0},\mathcal{P}^{1+J}_{1},\mathcal{P}^{2+J}_{2},\mathcal{P}^{3+J}_{3},\cdots$ with $n=m+J$. We shall use the $J=0$ sequence known as the diagonal sequence.

This method will be applied to our system of Boltzmann equations to extrapolate their solution from $z \ll1$, where the equations are derived analytically, to the $z\simeq 1$ case. It is understood that although above we considered a Taylor expansion about $z=0$ (which was assumed to be in the region of analyticity of $f(z)$), the discussion can be straightforwardly extended  for Taylor expansions about any other point inside the region of analyticity of $f(z)$. The application of Pad\'e approximants and justification of Pad\'e approximants are well described in~\cite{Bender}. 

We will conclude with an example which is related to the calculation of lepton asymmetry. Consider 
\begin{equation}
\label{ PadeExample}
f(z)=\left(0.0001
   z^{29/3}+0.0004
   z^{28/3}-0.0015
   z^{22/3}+0.0088
   z^{16/3}+0.0001
   z^{2/3}-0.0381 z\right)
\end{equation}
and Taylor expand $f(z)$ about $z=.7$. The corresponding $ \mathcal{P}^{7}_{7}(z)$ is 
\begin{equation}
\label{ PadeExample2}
 \mathcal{P}^{7}_{7}(z)=\frac{u_7(z)}{d_7(z)}
\end{equation}
where
\begin{multline*}
u_{7}(z)=-0.0253701 
   +0.00655229  
   (z-0.7)^7+0.00566604 
   (z-0.7)^6+0.00709687 
   (z-0.7)^5\\
   +0.020408  
   (z-0.7)^4+0.0213818 
   (z-0.7)^3\\
   +0.0358012 
   (z-0.7)^2-0.0161927 
   (z-0.7)
\end{multline*}
and
\begin{multline*}
d_7(z)=0.0242858
   (z-0.7)^7-0.0453282
   (z-0.7)^6+0.0524747
   (z-0.7)^5\\
   -0.0733029
   (z-0.7)^4-0.0319164
   (z-0.7)^3+0.232236
   (z-0.7)^2\\
   -0.502196
   (z-0.7)+1.0000000000000000.
\end{multline*}
The convergence of the diagonal sequence can be seen from $\mathcal{P}^{8}_{8}(1.44)=0.000465789$, $\mathcal{P}^{7}_{7}(1.44)=0.000463733$ and $ \mathcal{P}^{6}_{6}(1.44)=0.000449957$.

  \end{document}